\documentclass[aps,pre,10pt,twocolumn,noshowpacs,showkeys,amsfonts,amssymb,amsmath,longbibliography,superscriptaddress,notitlepage]{revtex4-2}
\usepackage[utf8]{inputenc}
\usepackage{graphicx}
\usepackage[caption=false]{subfig}
\usepackage{hyperref}

\usepackage[dvipsnames]{xcolor}

\begin{document}
%
%
%
%
\title{Predicting the diversity of early epidemic spread on networks}

\author{Andrea J. \surname{Allen}}
\affiliation{Vermont Complex Systems Center, University of Vermont, Burlington, Vermont}

\author{Mariah C. \surname{Boudreau}}
\affiliation{Vermont Complex Systems Center, University of Vermont, Burlington, Vermont}
\affiliation{Department of Mathematics \& Statistics, University of Vermont, Burlington, Vermont}

\author{Nicholas J. \surname{Roberts}}
\affiliation{Vermont Complex Systems Center, University of Vermont, Burlington, Vermont}

\author{Antoine \surname{Allard}}
\affiliation{D\'epartement de physique, de g\'enie physique et d'optique, Universit\'e Laval, Qu\'ebec (Qu\'ebec), Canada G1V~0A6}
\affiliation{Centre interdisciplinaire en mod\'elisation math\'ematique, Universit\'e Laval, Qu\'ebec (Qu\'ebec), Canada G1V~0A6}
\affiliation{Vermont Complex Systems Center, University of Vermont, Burlington, Vermont}

\author{Laurent \surname{H\'ebert-Dufresne}}
\affiliation{Vermont Complex Systems Center, University of Vermont, Burlington, Vermont}
\affiliation{Department of Mathematics \& Statistics, University of Vermont, Burlington, Vermont}
\affiliation{D\'epartement de physique, de g\'enie physique et d'optique, Universit\'e Laval, Qu\'ebec (Qu\'ebec), Canada G1V~0A6}
\affiliation{Department of Computer Science, University of Vermont, Burlington, Vermont}
\date{\today}
\begin{abstract}
The interplay of biological, social, structural and random factors makes disease forecasting extraordinarily complex.
The course of an epidemic exhibits average growth dynamics determined by features of the pathogen and the population, yet also features significant variability reflecting the stochastic nature of disease spread.
In this work, we reframe a stochastic branching process analysis in terms of probability generating functions and compare it to continuous time epidemic simulations on networks. 
In doing so, we predict the diversity of emerging epidemic courses on both homogeneous and heterogeneous networks. 
We show how the challenge of inferring the early course of an epidemic falls on the randomness of disease spread more so than on the heterogeneity of contact patterns. 
We provide an analysis which helps quantify, in real time, the probability that an epidemic goes supercritical or conversely, dies stochastically.
These probabilities are often assumed to be one and zero, respectively, if the basic reproduction number, or $R_0$, is greater than 1, ignoring the heterogeneity and randomness inherent to disease spread. 
This framework can give more insight into early epidemic spread by weighting standard deterministic models with likelihood to inform pandemic preparedness with probabilistic forecasts.
\end{abstract}
\maketitle
%
%
%
%
%
%
\section{Introduction}
%
By the time of this writing, the COVID-19 pandemic had reached every corner of the world. Public health efforts are now focused on identifying new clusters of outbreaks and their risk of causing new epidemic waves, much like they did at the beginning of the pandemic.  
As large outbreaks soared early on in a handful of countries, sporadic clusters of confirmed cases dotted regions in the United States.
Data surrounding new clusters or waves tend to consist of low numbers of cases highly sensitive to noise, sparking concern and uncertainty at the expected progression of the epidemic. 

The first confirmed case of COVID-19 in the US was reported on January 21st, 2020 in the state of Washington \cite{wash_firstcase_cdc}. Three subsequent cases were later identified in Washington; two hospitalizations on February 19th \cite{feb19_cases}, and two deaths on February 26th, one week later \cite{feb26_deaths_covid}. Then, on February 28th, a high school closed immediately after one of its students tested positive for a strain that had been associated with the January 21st case \cite{school_closed_news}. With limited knowledge of active cases, it was nearly impossible to predict the current and future severity of the outbreak.

One critical question in Washington after over a month with only a handful of detected cases, was whether this chain of events suggested a single tree of very few local transmissions, or multiple distinct introduction events from abroad. Despite decades of disease modeling, the community was ill-equipped to answer this question. 
The problem is challenging in part because of inadequate testing at the time, and also because well-established disease models often operate on deterministic mechanisms designed to describe the average behavior of large epidemics and not the random, discrete nature of small transmission chains. The looming question of whether a local COVID-19 outbreak would die off by itself or become a disaster, can only be modeled using tools capturing the stochasticity, or randomness, of person-to-person contact. To accurately model the potential outcomes of an epidemic based on limited case data, tools that capture the random nature of disease spread along with the structure of the population are required.

In this paper, we analyze the diversity of early epidemic courses. In doing so, we also hope to provide analytical tools to inform disease forecasts by accounting for the heterogeneity and stochastic nature of disease transmission.

Since the introduction of mean-field epidemic models, deterministic models of disease spread have continued to evolve in complexity and detail. Kermack and McKendrick's early work \cite{kermack1927contributionA, kermack1927contributionB, kermack1927contributionC} gave rise to compartmental models, in which the population under study is divided into two or more states. Perhaps the most widely known of these models is the susceptible-infectious-recovered (SIR) model, where the population is divided into susceptible, infectious, and recovered states (or compartments) and the trajectory of the sizes of each compartment can be tracked analytically over time \cite{AndersonMay, KeelingRohani}. 
The standard compartmental model assumes homogeneous mixing of the population and is deterministic, meaning that a given set of initial conditions and disease transmission rates always leads to the same expected outcome. A common extension to compartmental models is to relax the assumption of homogeneous mixing. One method for doing so is to derive mean-field equations for an epidemic process over contact networks, thereby introducing heterogeneous structure into the population \cite{pastor2015epidemic}. Similarly, it is possible to partition the population based on traits such as age, risk behaviors, or location and define how these partitions mix \cite{Ibana1990, Huang1992, Bolker1995, Lloyd2004}. While these approaches introduce more realistic contact behavior into a model, they fail to account for the inherently stochastic nature of disease spread; something of particular importance early in an outbreak.

Models based on stochastic processes address the shortcoming of deterministic outcomes in the standard mean field compartment models. A commonly used approach is that of branching processes. Bienayme-Galton-Watson processes are one widely used example, as they provide a good approximation of more general stochastic epidemic models \cite{Ball1995}. Beyond Bienayme-Galton-Watson processes, there exist a number of extensions such as including population structure, multiple types of hosts/pathogens, and considering time to be continuous rather than discrete \cite{Ball1997, L_Allen2015}. In these branching process models the basic reproduction number, \(R_0\), the probability of an outbreak, and the final proportion of population infected (in a ``supercritical" model) are typically tractable to compute. While these are all important, a shortcoming of most branching models is the difficulty of tracking the trajectory of outbreaks through time and knowing whether it matches the continuous time dynamics of real epidemics. Stochastic differential equations are an alternative modeling approach that allow one to track outbreak trajectories, as well as often finding threshold conditions for the occurrence of an outbreak or the existence of an endemic equilibrium \cite{LAllen2017, Wang2018, Gray2011}. Like all models, stochastic differential equations have drawbacks; the most relevant is standard formulations do not allow for stochastic extinction if \(R_0 > 1\).

Another common approach in disease modeling is times series analysis, more statistical in nature than mechanistic models. This theory can be applied to assist in estimating the parameters of compartmental models or to combine ensembles of compartmental models to increase prediction accuracy
\cite{Finkenstadt2000, Zhan2018}. Independently of compartmental models, time series analysis can be used to study covariates of disease occurrence (\emph{e.g.}\ weather), estimate the future variability in observed cases, or to make epidemic forecasts \cite{Allard1998, Lopman2009, Hu2006}. A necessary requirement for the effective use of many time series methods however is data. When facing sparse incidence numbers, and in the absence of historical data, the methods become problematic and thus are not suitable for emerging diseases.

Agent-based models are another family of models used for tracking epidemic progression, in which agents, or individuals in the population, are tracked throughout the course of the epidemic. Agents are parameterized with individual attributes, capturing the heterogeneity of the population and aspects from compartmental models are used to categorize the state of each agent \cite{silva2020covid, gharakhanlou2020spatio}. While there is great power in adjusting various attributes for different epidemic conditions and environmental factors, most of these models are computationally expensive and need a copious amount of information to generate the entire collection of agents \cite{cuevas2020agent, hoertel2020facing, silva2020covid, staffini2021agent, gharakhanlou2020spatio, srikrishnan2021small}, making them ill-suited for modeling early epidemic spread with a handful of cumulative case counts and sparsely available data.

Early in an outbreak, we often face the unique challenge of modeling disease spread while taking into account the heterogeneity of the population and the stochastic nature of disease spread, including stochastic extinction, without substantial amounts of data. The heterogeneous contact structure found in populations is accounted for by network models, and a first approximation for a relevant contact structures in a novel outbreak can be taken from past outbreaks of similar diseases. Including a sufficient number of possible states will typically account for heterogeneity in host and pathogen type. The randomness of transmission is modeled with stochastic processes, many of which easily permit stochastic extinction.

The above considerations naturally lead to percolation theory, which can be used to analyze stochastic compartmental disease models on networks. Percolation models unite contact heterogeneity and stochasticity under a single modeling framework \cite{meyers2007contact}. 
An underlying contact network acts as the substrate for disease to propagate through, resulting in a directed network of transmission \cite{kenah2007second, miller2007epidemic, kenah2011epidemic}. The resulting epidemic percolation networks can be analyzed using branching process theory \cite{AthreyaNey, newman2001random} which model stochastic transmission between individuals using an underlying offspring distribution. Branching processes are especially useful for early epidemic modeling, as they allow for stochastic behavior of spread as well as stochastic extinction \cite{Miller2018bp}.
Specifically, the method of probability generating functions (PGFs) can be used to analyze branching processes on percolation networks \cite{newman2001random, newman2002spread, Miller2018bp}.
Consequently, there have been many recent applications of this framework designed specifically for COVID-19 \cite{levesque2021model, bertozzi2020challenges, mitrofani2021branching, zhang2021heterogeneous, akian2020probabilistic, kojaku2021effectiveness}.

The PGF formalism is traditionally used for estimating quantities that pertain to the predicted end of an epidemic --- such as the probability of infecting a macroscopic fraction of the population and distribution of final outbreak sizes --- but not how risk and outbreak sizes change dynamically over time. 
Kenah and Robins show how modified percolation models (epidemic percolation networks) have a final state isomorphic to a network-based SIR models \cite{kenah2007second}.
Most bond percolation frameworks differ from SIR dynamics as SIR transmission events are correlated through the distribution of the infectious period of each infected individual whereas percolation models assume independent contacts and transmission events.
More importantly, percolation models integrate over time to map transmission dynamics (which occur in continuous time) to discrete bond percolation (which occur in discrete time with a fixed probability of transmission).

In 2009, No\"el \textit{et al.}~\cite{noel2009time} offered a novel method for tracking the stochasticity of outbreak sizes by epidemic generations, allowing us to incorporate discrete time into the percolation-framework model. In this paper, we show how the generation-based PGF formalism also succeeds in tracking emerging epidemic size in \textit{continuous} time, by validating the PGF approach with event-driven simulations on networks. This result allows us to use PGFs and early disease data to quantify epidemic risk and survival probability.


%
%
%
%
\section{Theoretical Analysis and Simulations\label{sec:methods}}
%

\subsection{Probability generating functions\label{sec:pgfFormalism}}
\begin{figure}[t]
    \centering
    \includegraphics[width=\linewidth]{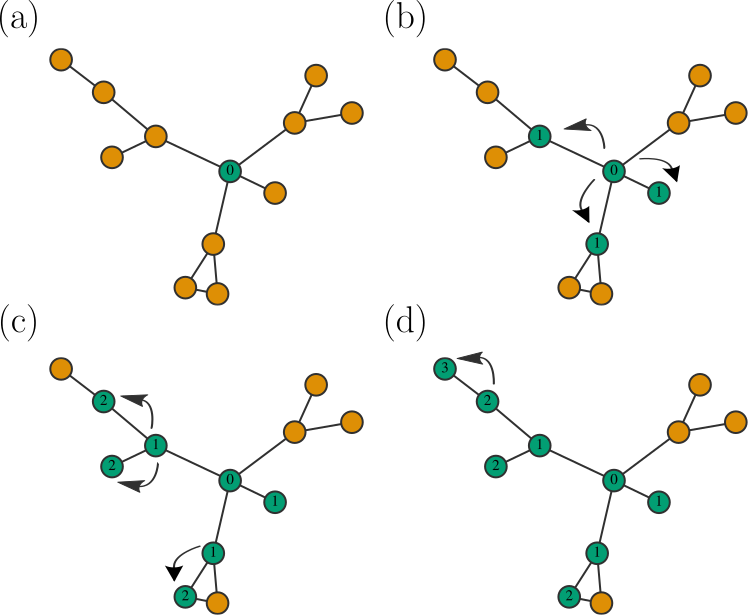}
    \caption{\textbf{Schematic of generations of infection through a network.} Each node's label corresponds to the epidemic generation in which it was infected. The initial infected node is in generation 0, any nodes they infect constitute generation 1, and so on.}
    \label{fig:cartoon_network}
\end{figure}

PGFs succinctly encode a probability distribution in a power series representation so that the methods of power series analysis can be applied \cite{wilf2005generatingfunctionology}. PGF theory naturally extends to disease modeling, where the distribution under study encapsulates a disease transmission network, framed as a bond percolation problem where the bond occupation probability $T$ is the probability of an infected individual infecting one of their contacts over the course of the entire epidemic \cite{newman2001random, newman2002spread}. Typically, this approach is used to solve for the average behavior of the system; we can solve for quantities such as the critical transmissibility at which the entire connected population will become infected, or the distribution of outbreak sizes. However, an increasing necessity of disease modeling is to model early epidemic spread, analyzing early cases to predict whether an outbreak will become large before it actually happens. In 2009, No\"el \textit{et al.}~\cite{noel2009time} developed the epidemic PGF modeling theory further to model the sizes of progressive epidemic \textit{generations}, demonstrated in Fig. \ref{fig:cartoon_network}.

The foundations for both aforementioned generating function methodologies are the same, beginning with the underlying contact network. In a contact network, we represent a collection of individuals as \textit{nodes} and their contacts between each other with \textit{edges}. We say that two nodes are \textit{neighbors} if they are in contact, i.e. connected by an edge. A node's \textit{degree} is how many neighbors it has. The \textit{degree distribution} of a network is the probability distribution for the number of neighbors of one node. Under an SIR disease modeling framework, nodes begin as \textit{susceptible}, and become \textit{infectious} if it is infected by one of its neighbors, which occurs with probability $T$.

The framework introduced by No\"el \textit{et al.}~uses PGFs to describe generations of infection as a piece-wise generating function, which can then be studied using branching process techniques. First we introduce what an epidemic \textit{generation} is. We say a node belongs to generation $g$ if it became infected via a neighbor belonging to generation $g-1$.
Assuming an infinite-size random network drawn from a specific degree distribution (a process known as the configuration model \cite{FosdickBaileyK2018CRGM}), each chain of infections stemming from an initial infected case, \emph{patient zero}, can be considered uncorrelated. This uncorrelated assumption follows from configuration models having locally treelike structure, thus every  subsequent case to be treated as a node that was reached by following a random edge. In this way, each node in each generation can be treated as independent from all other nodes in its generation. Thus, for each node in generation \(g\), the PGF describing the distribution of cases that node will cause over the course of the epidemic is given by
\begin{equation}
    G_{g}(x; T) = 
    \begin{cases}
    G_{0}(x; T) & (g=0)\\
    G_{1}(x; T) & (g > 0)\\
    \end{cases}
    \label{eq:generation}
\end{equation}
where $G_g(x;T)$ is the distribution, in PGF notation, of the \textit{secondary} cases caused by a single node in generation $g$. Now, we will provide the derivations used to obtain this framework using the underlying network, generating functions, and branching process theory.

Using PGF notation, we will refer to the original underlying network degree distribution as $G_0(x)$, which we write as
\begin{equation}
    G_{0}(x) = \sum^{\infty}_{k=0} p_{k}x^{k}. \label{eq:g0}
\end{equation}
The $k$th coefficient of Eq. \eqref{eq:g0}, $p_{k}$, is the probability of randomly choosing a node with degree $k$ from the network. The average degree of the network is denoted as $\langle k \rangle$, derived by the first derivative of the generating function as
\begin{equation}
     G'_{0}(1) = \langle k \rangle = \sum^{\infty}_{k=0} kp_{k}.
\end{equation}
To study the progression of an epidemic, we are interested in the distribution of infections from each subsequently infected node. Before introducing transmission probability, we work first with the aforementioned degree distribution to understand how many infections each node could cause through each generation. Assuming an initial infectious node, patient zero, we know $G_0(x)$ is the distribution of contacts for them, but that distribution is different for anyone patient zero infects. This phenomenon is known as the friendship paradox; the degree of a node chosen by following a random edge is on average, larger than the degree of the node selected at random whose edge we followed. In this context, patient zero has a degree distribution of $G_0(x)$, but the node who patient zero first infects has a degree distribution known as the \textit{excess degree distribution}, denoted as $G_1(x)$ in PGF notation. To obtain $G_1(x)$, we are interested in the degree of nodes provided that we arrive there by following the edge from one of its neighbors. So, this means the resulting distribution will exclude that neighbor, reducing every node's degree by 1, and multiplied by the number of ways they could have been reached, which is the original degree. This algorithm surmounts to taking the derivative of $G_0(x)$, so that we have the excess degree distribution
\begin{equation}
     G_{1}(x) = \frac{\sum_{k}(k+1)p_{k+1}x^{k}}{\sum_{k}(k+1)p_{k+1}} =  \sum^{\infty}_{k=0} q_{k}x^{k} \label{eq:G1}
\end{equation}
and where the derivative is divided by the average degree of the network $\langle k \rangle$ in order to normalize the distribution tuned to the original node.
The coefficients $q_{k}$ represent the probability of reaching a node with degree $k$ from a randomly chosen edge. 

Returning to the percolation problem, we incorporate disease transmissibility $T$ to transform the excess degree distribution into a \textit{secondary case distribution}. The probability that a single infectious node infects $l$ neighbors given it has degree $k$, or $k$ neighbors, is given by
\begin{equation}
    p_{l|k} = \binom{k}{l}T^{l}(1-T)^{k-l}
\end{equation}
From this we can derive the PGF for the number of infections caused by patient zero, which we denote $G_0(x;T)$ for short, given by
\begin{align}
    G_0(x;T) & = \sum_{l = 0}^{\infty}\sum_{k=l}^{\infty} p_{k}p_{l|k}x^{l} \nonumber \\
      & = \sum_{k = 0}^{\infty}\sum_{l=0}^{k} p_{k}\binom{k}{l}T^{l}(1-T)^{k - l}x^{l} \label{eq:offspring_dist} \nonumber \\
      & = G_{0}(1 + (x-1)T).
\end{align}
From $G_{0}(x;T)$, $ G_{1}(x;T)$ can be calculated in a parallel fashion as $G_1(x)$ is from $G_0(x)$. The PGF $ G_{1}(x;T)$ is now the probability distribution of the number of infections caused by a single node, i.e., the secondary case distribution.

We now present how to study the evolution of the distribution of cumulative cases for the percolation model following No\"el \emph{et al}. Let $s$ be the number of cumulative cases at generation $g$ and let $m$ be the number of infectious nodes strictly belonging to generation $g$. (Note that in this way, $s$ is the sum of all $m$ values from generation 0 up to and including generation $g$.) We let the probability of having \emph{s} total infections by the end of the \emph{g}-th generation with \emph{m} becoming infected (and thus being infectious) during that generation be denoted as $\psi_{sm}^{g}$ \cite{noel2009time}. This has an associated probability generating function, given by
\begin{align}
    \Psi_{0}^{g}(x,y) = \sum_{s,m} \psi_{sm}^{g}x^{s}y^{m}
\end{align}
over all $s, m$. 

We know the distribution of infections following from a single infectious node in generation \({g-1}\) is generated by \(G_{g-1}(1+(x-1)T)\) (from Eq. \eqref{eq:offspring_dist}). The PGF of a finite sum of independent processes is the product of their PGFs, and as discussed above, each node in generation \({g-1}\) can be treated independently. Thus, if we assume the state in generation \({g-1}\) is given by the pair \((s',m')\), then the probability of spawning \(m\) new infectious nodes in generation \(g\) is generated by
\begin{align}
    \sum_{m} P(m|s', m')x^{m} = [G_{g-1}(x;T)]^{m'}
\end{align}
where the equality occurs as a result of the right side describing the probability of \(m\) infectious nodes in generation \(g\) assuming \(m'\) such nodes at \(g-1\) from branching process theory.

For a given state \((s',m')\) in generation \(g-1\), \(m\) new infections will result in $s'+m$ cumulative infections in generation $g$. So, having $m$ new infections occurs with probability \(\psi^{g-1}_{s'm'}P(m|s', m')\), where the \(\psi^{g-1}_{s'm'}\) term is the probability of being in the state $(s',m')$ at generation $g-1$. Now, we can re-write the entire PGF for the state space of $(s,m)$ at generation $g$ as
\begin{align}\label{eqn:psi_derived}
    \Psi_0^g (x,y) =& \sum_{s,m} \psi_{sm}^g x^s y^m  = \sum_{s',m} \psi_{sm}^g x^{s'} (xy)^m \\
    \nonumber =& \sum_{s'm'}x^{s'}\sum_{m}\psi_{s'm'}^{g-1}P(m|s', m')(xy)^{m} \\
    \nonumber =& \sum_{s',m'} \psi^{g-1}_{s'm'} x^{s'} \sum_m P(m|s',m')(xy)^m \\
    \nonumber =& \sum_{s'm'}\psi_{s'm'}^{g-1}x^{s'} [G_{g-1}(xy;T)]^{m'}\\
    =& \Psi^{g-1}_0 (x, G_{g-1}(xy;T))
\end{align}
This defines a recurrence relation when \(g\geq1\). Taking \(\Psi_0^0 = xy\) as the assumption that there is only one initial infectious individual, then \(\psi^0_{sm}=\delta_{s1}\delta_{m1}\).

\begin{figure}[t]
    \centering
\includegraphics[width=0.9\linewidth]{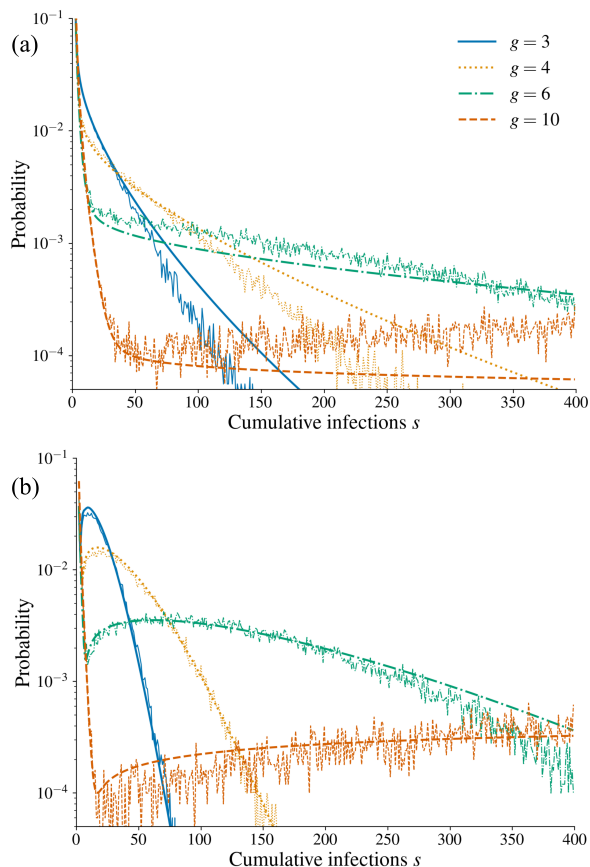}
    \caption{\textbf{Time evolution of epidemics on homogeneous and heterogeneous networks.} We show the probability of having $s$ cumulative cases by and including generation $g$ for select generations (Eq. \ref{eqn:prob_s_g}). Panel (a) shows the results on a modified power-law random networks with degree distribution given by $p_k = k^{-2}e^{-k/10}$ with average degree $\langle k \rangle = 1.79$, average excess degree $\langle q \rangle = 3.04$, $\beta = 0.004$ and $\gamma = 0.001$ such that $R_0 = T\frac{\langle k^2 \rangle - \langle k \rangle}{\langle k \rangle} = \frac{\beta}{\beta + \gamma}\frac{\langle k^2 \rangle - \langle k \rangle}{\langle k \rangle}= 2.44$. The smooth lines show the theoretical prediction for the probability distribution of cumulative infections. The distributions are validated by 75,000 simulations performed on 150 random network realizations with 10,000 nodes, following the process outlined in Sec. \ref{sec:simulations}.
    Panel (b) shows the results of equivalent analysis and simulations on Erd\H{o}s-R\'{e}nyi random networks with $\langle k \rangle=2.5$, $\beta = 0.004$ and $\gamma = 0.001$ such that $R_0 = 2.0$.}
    \label{fig:time}
\end{figure}

Our primary focus in this paper will be on the distribution of cumulative infections $s$ in each generation $g$. We derive a generating function for this quantity by taking the marginal distribution over $y$ of Eq.~ (\ref{eqn:psi_derived}). We let the coefficient $p_s^g$ be defined as the probability of having $s$ cumulative cases at generation $g$. To derive $p_s^g$, we wish to take the sum over all values of $m$ for which the state $s,m$ holds at generation $g$. To do so, we set the counting variable $y$ of new cases simply equal to 1. As such, the coefficients $p_s^g$ are generated by
\begin{align}\label{eqn:prob_s_g}
    \Psi_0^g(x, 1) = \sum_{s, m} \psi_{sm}^g x^s = \sum_s \sum_m \psi_{sm}^gx^s = \sum_s p_s^gx^s.
\end{align}
Now the generating function in Eq.~(\ref{eqn:prob_s_g}) defines a probability distribution over $s$ for each generation $g$, and is our main quantity under study. The analytical distributions are illustrated in Fig. \ref{fig:time} along with event-driven simulations to validate the theory.

\subsection{Simulations of continuous SIR dynamics}\label{sec:simulations}

For a realistic model of the spread of disease in a population, we simulate a stochastic disease process of an SIR epidemic on synthetic contact networks in continuous time \cite{andrea_allen_2021_5076514}. We use an event-driven framework, which is advantageous for epidemic modeling, because it is much faster compared to a brute-force time-step simulation due to its leveraging of the Markovian dynamics of infectious and recovery periods of individuals \cite{kissmathonnetworks2019, miller_ting_2019, bauer_engblom_widgren_2016}.
Recall in the SIR model that nodes inhabit the susceptible, infectious, and recovered states as the disease progresses, where nodes become infected if one of their infectious neighbors transmits to them. The standard SIR model is governed by two rate parameters; $\beta$, the rate per unit time of an infectious node transmitting to other nodes, and $\gamma$, the rate per unit time of an infected node recovering.
In a continuous time event-driven simulation, infection and recovery are Poisson processes occurring at rates \(\beta\) and \(\gamma\) respectively, and relate back to the percolation framework by defining transmissibility \(T=\beta/(\beta + \gamma)\).

We draw a random network from a given degree distribution, and begin the simulation algorithm by assuming a random initial infectious node, patient zero, with degree $k_0$. Patient zero could either recover before transmitting to any of its neighbors, or infect one or more of its neighbor nodes. The stochastic process governing the behavior of a single infected node is the superposition of $\hat{k}+1$ Poisson processes, where $\hat{k}$ is the number of susceptible neighbors, and with one extra process governing the time until recovery. Say patient zero infects Neighbor 1, who has $k_1$ neighbors. Then with two infectious nodes, the stochastic process encompassing all possible events is a Poisson process with rate \((\hat k_0 -1) \beta + \hat k_1 \beta + 2 \gamma\), and so on as more nodes become infected.

 Each possible event given by the sub-processes is the first to occur with probability \(i/(\hat k \beta + \gamma)\) where \(i \in \{\beta, \gamma\}\), with the Poisson process rate term from $\hat{k}$ reducing if an infection event occurs, and stopping entirely if the contagious node recovers. The disease process for the whole population is a natural extension of that described above, with each node assumed identical apart from degree. The evolution of the unmitigated disease process from here is intuitive, either eventually all the infectious nodes recover or the whole connected population becomes infected.

Computationally, the above process is simulated by generating a random network from a given degree distribution using a large enough number of nodes, $N$, such that average degree $k \ll N$. As we cannot simulate numerically on an infinite network, the best choice for $N$ is the largest value the numeric simulation can support. A node is randomly selected to be patient zero, and the disease spread proceeds via stochastic event-driven simulation, often known as the Gillespie algorithm \cite{gillespie1977}.  Continuous time is tracked using a random variable $\tau$, known as the waiting time, which is exponentially distributed with parameter the sum of the rates of all the potential infection and recovery events. Each competing process is the first to occur with probability of its own rate divided by the sum of all rates of that process type, as described by the Poisson process above. The simulation is advanced via this algorithm until either there are no more infectious nodes or until there are no more susceptible nodes, and allows for obtaining the resulting evolution of the disease spread in terms of both generations of infection and continuous time.

\section{Results\label{sec:results}}

We employ the generational size distribution theory to explore the evolution of epidemic size on a variety of network structures, and compare the generating function theory against continuous-time simulations. We use the event-driven simulation framework so that we can track the progression of the epidemic in both continuous time as well as the generation sizes corresponding with the branching process, which allows us to validate the theoretical distributions, as well as introduce a preliminary prediction for the expected continuous time emergence of successive generations. Then, we use the PGF framework to measure the probability of an epidemic surviving, or continuing on, past an arbitrary generation, depending on the characteristics of the network and disease.

\subsection{Time evolution on homogeneous and heterogeneous networks}\label{sec:epi_prob}
In Fig. \ref{fig:time} we show the probability distributions of cumulative infections by the specific generation for two network models. It is noteworthy that this modeling method holds for configuration model networks with varying types of degree distributions. Here, we show the results on a modified power law network and an Erd\H{o}s-R\'{e}nyi (ER) network both used in Ref. \cite{noel2009time}. The ER network has mean degree and excess degree $\langle k \rangle = \langle q \rangle = 2.5$, while the modified power law has mean degree $\langle k \rangle = 1.79$ and average excess degree $\langle q \rangle = 3.04$, a more heterogeneous distribution. We demonstrate that the distributions of outbreak size appear to be more a result of the stochastic nature of the disease spread, rather than the structure of the network, though the structure does play a role in the shape of the distribution. 

Our results convey that there is not one clear trajectory of a typical large outbreak, in contrast to traditional results with deterministic modeling.  Instead, the stochastic nature of epidemic size is captured by a long tail in the distribution of cumulative cases over each epidemic generation. One unique aspect of this paper is that we validate this result using continuous-time simulations showing the same shape and long tail in outbreak size distributions as our analytical results. We do anticipate the simulated distributions and analytical distributions to vary from each other due to a few factors including the finite-size effects of simulated networks, and the fact that we compare a discrete analysis with a continuous-time process, but the general behavior appears consistent throughout the different generations. 

We also find that on both the heterogeneous network and the homogeneous network, there is a high probability of an outbreak going extinct before growing large, however, if it does take off, the distribution levels off over the space of epidemic size. That is to say, if indeed an epidemic takes off and has arrived at generation six, via a transmission chain of length six, there is an almost equal probability of having anywhere from 50 to 500 cumulative cases by the time generation six is reached. We emphasize that these results display the unpredictability in early stages of epidemics, even ignoring the difficulty of estimating model parameters, it is near impossible to infer with much confidence how many infections there may actually be in the population.

\begin{figure}[t]
        \centering
        \includegraphics[width=\linewidth]{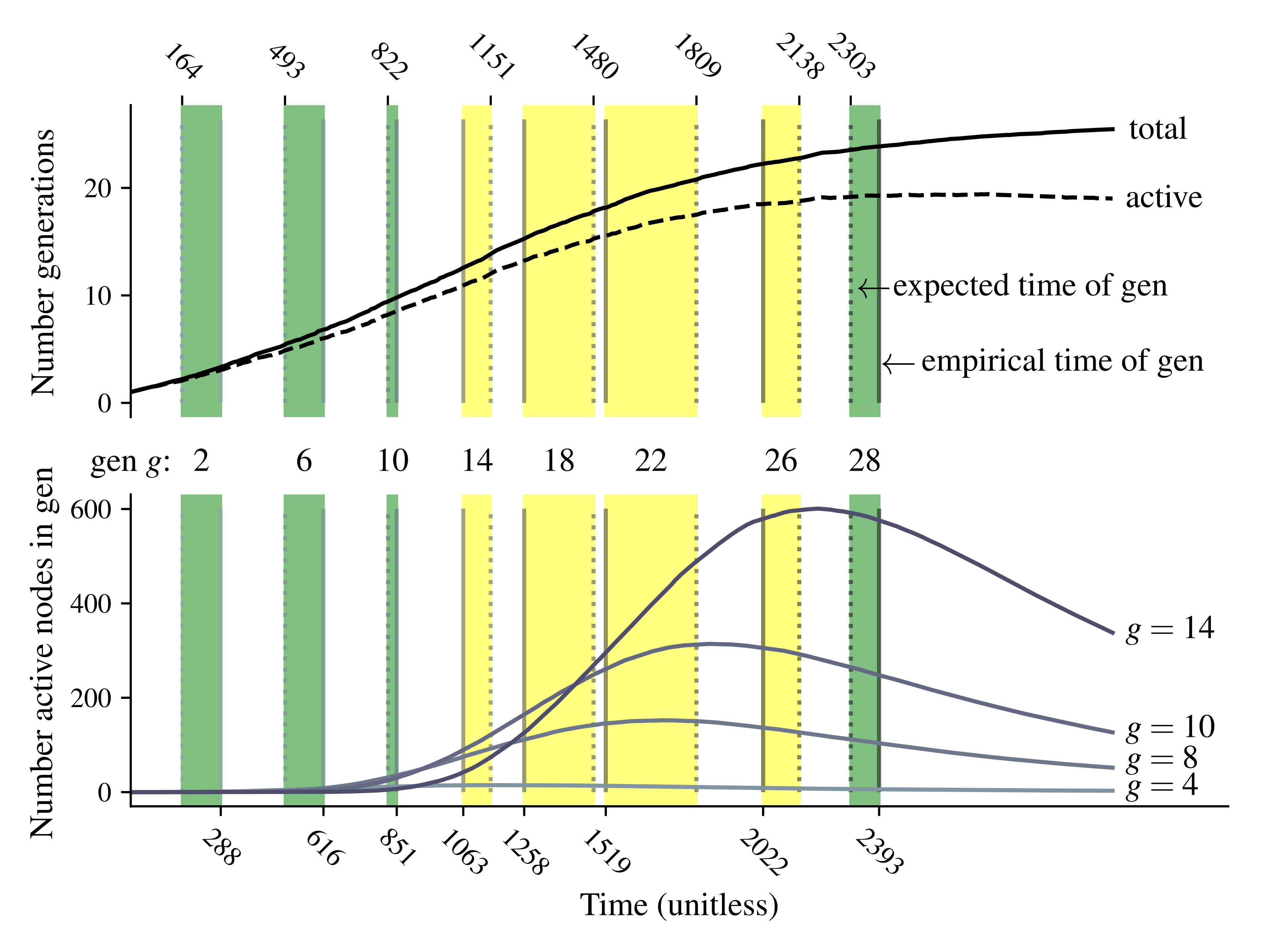} 
\caption{\textbf{Time evolution of the active epidemic generations and emergence times.} Top panel (curves): average number of total and \textit{active} generations at time $t$ for the modified power-law network with degree distribution $p_k = k^{-2}e^{-k/10}$.
Bottom panel (curves): average number of active \textit{nodes} belonging to each generation shown over time to accompany the top panel. The tick marks in the top panel (and dotted vertical lines) correspond to increments of $g/\langle q \rangle \beta$, the predicted generational emergence times, and the bottom tick marks (and solid vertical lines) correspond to the average empirical time at which that generation $g$ emerged, for an example network. If the average time of emergence was greater than its respective $g/\langle q \rangle \beta$ value, that is, after the predicted time, the difference is highlighted in green. If the average empirical time was less than that predicted, the difference is highlighted in yellow.} \label{fig:timing1}
\end{figure}

\subsection{Generations of infection in continuous time}\label{sec:time_tracking}

While the behavior of the epidemic in our formalism is described by generations of infection, most applications of disease models desire descriptions of the dynamics in continuous time. We find early agreement from our model of generational infections with a distribution in continuous time, described in terms of the expected time of emergence of an arbitrary generation $g$. The agreement is surprising since one might not expect a consistent relationship between a generation number and the expected time of its emergence given the observed heterogeneity of early spread in Fig 2.
Yet, by defining the \textit{emergence} of generation $g$ as the time its first member is infected, we find a simple linear relationship that allows us to map the PGF framework to continuous time.

We can show that the expected time of emergence of an arbitrary generation $g$ is given by $$\mathbb{E}[t(g)] = \frac{g}{\langle q \rangle \beta}$$ where $\langle q \rangle = G_1^\prime(1)$  is the average excess degree of the network. We arrive at this expression for $\mathbb{E}[t(g)]$ via a simple argument over the Poisson process governing how nodes in generation $g-1$ can lead to the first cases of generation $g$. Each node of generation $g-1$ can recover at rate $\gamma$ but also has on average $\langle q \rangle$ neighbors they can infect at rate $\beta$. Therefore, the first event around them will occur at a combined rate $\alpha = \langle q\rangle \beta+\gamma$ and will lead to a case in generation $g$ with probability $T_q = \langle q \rangle\beta/(\langle q \rangle \beta+\gamma)$. The first infectious node in generation $g-1$ can therefore lead to the emergence of generation $g$ after $1/\alpha$ with probability $T_q$; if not, or the second node in generation $g-1$ could lead to the emergence of generation $g$ with probability $T_q(1-T_q)$ after $2/\alpha$ (approximate delay between the first and second node of generation $g-1$ plus the expected time to generation $g$); and so on for the third node and beyond. This sequence of possibilities can be summarized by an arithmetico-geometric sum,
\begin{align}
\mathbb{E}[t(g)-t(g-1)] & = \frac{T_q}{\alpha} \sum_{k=1}^\infty  (1-T_q)^{k-1}k \nonumber \\
& = \frac{T_q}{\alpha} \frac{1}{T_q^2} = \frac{1}{\langle q\rangle\beta} \; .
\end{align}

In Fig. \ref{fig:timing1}, we demonstrate in practice how the expected time of emergence of consecutive generations falls in line with the predicted time measure. To show intuitively why we see this phenomenon, we show the time evolution of the active epidemic generations. We track time in two ways; in continuous time following the event-driven process discussed in Sec. \ref{sec:simulations}, and also in terms of the expected time of emergence of each generation $g$, in the form $t = g/\langle q \rangle \beta$. 
We define a generation to be \textit{active} if it contains one or more nodes who are not recovered and have susceptible neighbors at time $t$ in the simulation. We illustrate the number of total and active generations over time, as well as the number of active \textit{nodes} belonging to each generation, which helps clarify the roles each generation plays in causing the next wave of infection over a given interval in continuous time. 

Having an understanding of the time at which a generation will emerge acts as a complement to the probabilities of extinction and cumulative cases discussed in Sec. \ref{sec:epi_prob} and \ref{sec:extinction_prob}. Equipped with the distributions describing the stochasticity of outbreaks, the expected time mapping can be a tool for analysis of the dynamics of the worst-case scenarios when an outbreak does occur.

\begin{figure}[t]
    \centering
    \includegraphics[width=\linewidth]{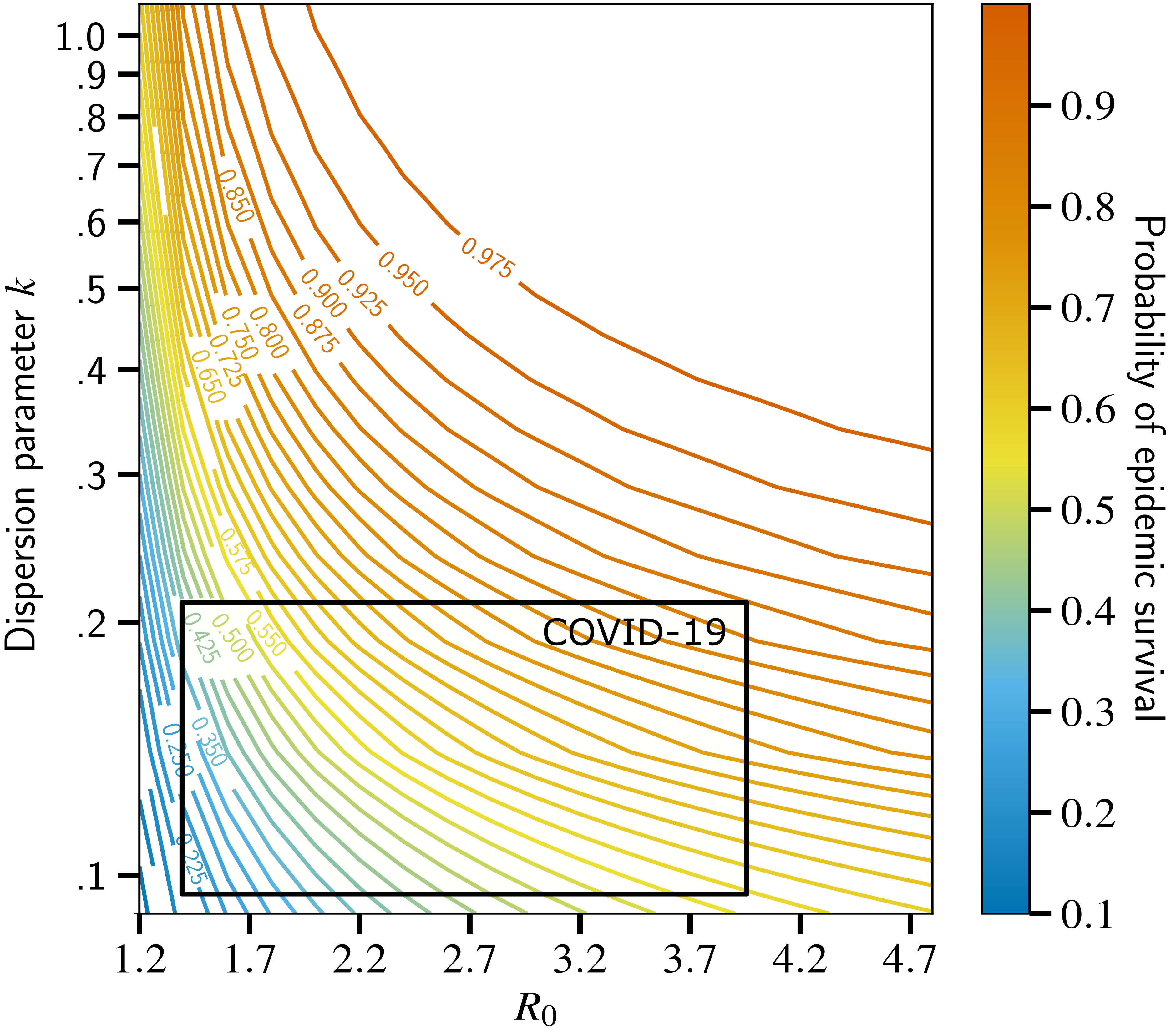}
    \caption{\textbf{Probability of epidemic survival as a function of contact structure.} 
    The contour plot shows the initial probability of epidemic survival for negative binomial distributions of infections over a range of possible $R_0$ values (average transmissions per case) and dispersion parameter $k$ (inverse of heterogeneity).
    The box highlights estimates for COVID-19 based on data from Wuhan, China \cite{hebert2020beyond}.
    We assume an epidemic generation of $g=4$ and $s=16$ cases which corresponds to the epidemic growing from 1 case to 16 over 4 generations. 
    Using a serial interval of 4 days, the average of the estimated range for COVID-19 \cite{du2020serial}, this tracks to roughly over two weeks of spread.
    Similarly, in the state of Washington, the first recorded case of COVID-19 occurred on January 21st, 2020 but following cases were only identified on February 19th and increased to 18 by March 2nd.
    This figure illustrates how these cumulative case data could have been used in real time with our theoretical tools to estimate epidemic risk.}
    \label{fig:contour_prob}
\end{figure}

\subsection{Probability of pandemics or stochastic extinction}\label{sec:extinction_prob}
The PGF generational theory can also be used to measure the probability that an emerging epidemic has a chance of dying off on its own, or ``surviving".
Deterministic models always predict that an epidemic will occur if \(R_0 >1\), that is, if the average number of secondary infections caused by an infectious individual is more than one. In reality, there is a non-zero chance the outbreak will die off by chance, shown in Fig. \ref{fig:contour_prob}. Branching process models have been used in theoretical epidemiology for estimating such probabilities \cite{Becker77, Becker74, DHB_book}. However, simple branching process models are Markovian in the number of active infections, \(m\). This is problematic in an applied setting as cumulative cases, \(s\), are often the available data. Moreover, we show that conditioned on reaching generation \(g\), the probability of the outbreak going extinct after generation \(g\) rather than becoming an epidemic is path dependent in the sense that the value of \(s\) at \(g\) changes the extinction probability, shown in Fig.~\ref{fig:covid_heatmap}.

To utilize the extinction probabilities, we want to look specifically at the variable \(\rho^g_s\), the probability that given $s$ cumulative cases at generation $g$ the epidemic will go extinct, or die off, sometime afterwards. Given that the evolution of \(m\) occurs as a branching process with the offspring PGF given by Eq. \eqref{eq:offspring_dist}, one can easily compute the probability of extinction of a single infection chain, \(p_e\), as the solution of $p_e = G_1(p_e;T)$ using branching process theory \cite{Miller2018bp}. The distribution of probabilities of \textit{reaching}  \((s,m)\) in the state space for each \(g\) is given by \(\psi^g_{sm}\), as discussed in Section \ref{sec:pgfFormalism}. We define a new distribution, that of the probability of the outbreak still being in existence in generation \(g\), by
\begin{align}
    \tilde \psi^g_{sm} = 
       \begin{cases}
          \displaystyle\frac{\psi^g_{sm}}{\displaystyle\sum_{s',m'>0} \psi^g_{s'm'}} & m > 0 \\
          \qquad 0 & \text{ otherwise }
       \end{cases}.
\end{align}
Thus, \(\rho^g_s\), the probability of the epidemic going extinct given it has arrived at $s$ cases by generation $g$ is given by
\begin{align}
    \rho_s^g = \sum_m \frac{\tilde \psi^g_{sm}}{\sum_{m'} \tilde \psi^g_{sm'}} p_e^m.
\end{align}
The probability of epidemic \textit{survival} for an epidemic being active in generation \(g\) with \(s\) cumulative infections is then given by \(1-\rho_s^g\). We illustrate an example of how the survival probabilities change depending on the underlying network and disease parameters in Fig.~\ref{fig:contour_prob}.

\subsection{Epidemic probability and COVID-19 data}\label{sec:covid}

We now apply the epidemic survival probability theory to early incidence of COVID-19 cases in the US.
This allows us to look at the evolution over time of public health risk, while taking into account the stochastic elements of the early spread. 
We assume a distribution of secondary infections parameterized as a negative binomial with $R_0$, the basic reproductive number, and $k$, the dispersion parameter of the contact network \cite{hebert2020beyond}. Together, these parameters determine the average behavior of disease spread where $k$ is responsible for the variation in secondary cases, in turn affecting the likelihood of superspreading events \cite{lloyd-smith_schreiber_kopp_getz_2005, althaus_2015, kucharski_althaus_2015}. 
A low dispersion parameter $k$ (high heterogeneity) means that a select few cases may cause the majority of secondary infections \cite{riou_althaus_2020}, which in our framework here might correspond to a single case leading to an extreme increase in cases in the next generation.
For that reason, it is often assumed that the early spread of an epidemic is highly sensitive to superspreading events \cite{althouse2020superspreading}. 
Yet, as shown in Fig. 2, heterogeneity in contact structure actually has less of an impact on the distribution of outcomes than the inherent stochasticity of transmission.

\begin{figure}[t]
    \centering
    \includegraphics[width=\linewidth]{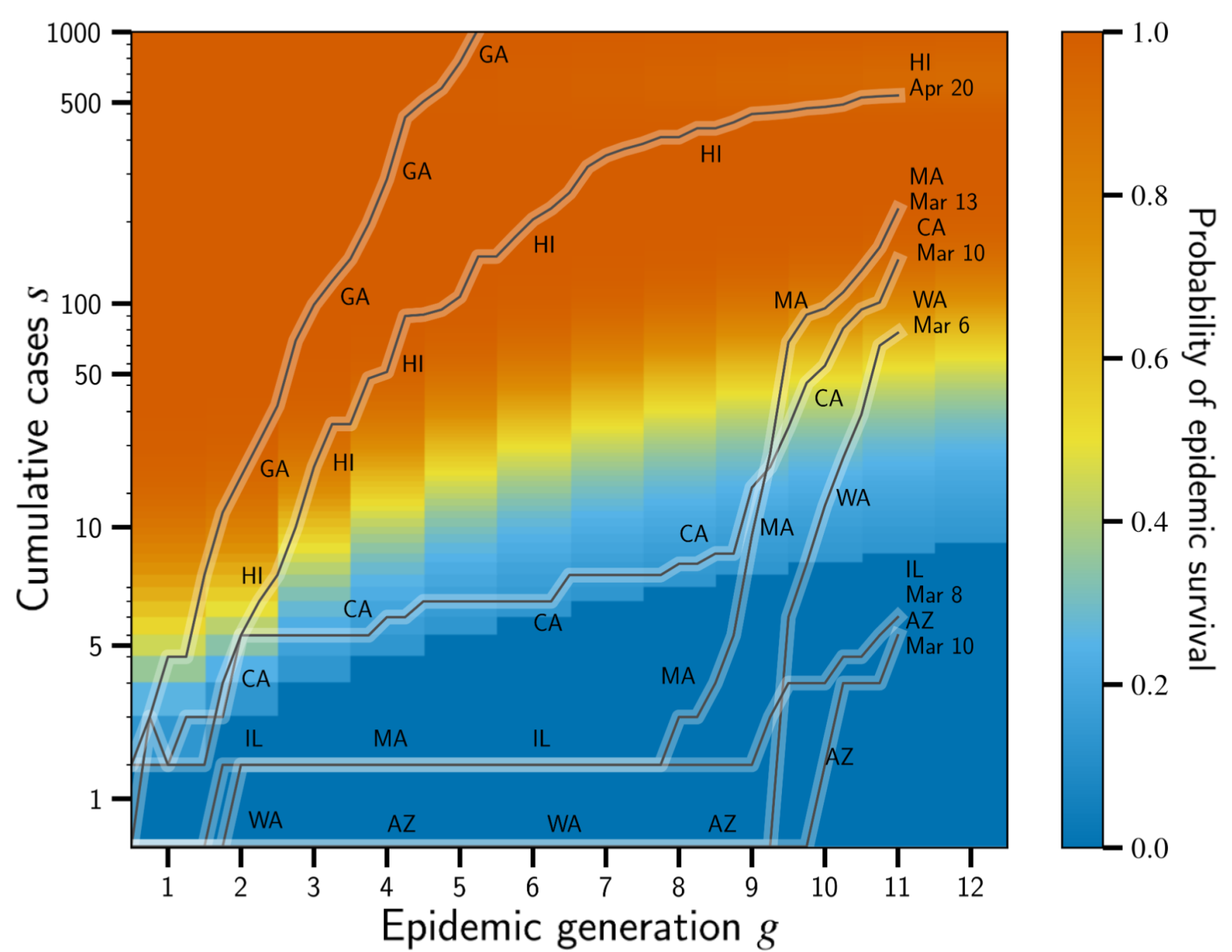}
    \caption{\textbf{Probability of epidemic continuing on as a function of case counts and time.} As a simple comparison, we use early data from the COVID-19 pandemic and show a selection of U.S. states following unique timelines from the first recorded case onward. This simple visualization is not meant as a validation but only to explore how quickly our predictions for the probability of the epidemic not dying off changes as an epidemic grows.
    To calculate these probabilities, we use a negative binomial distribution of secondary infections with $k=10^{-1}$, $R_0=2.5$ along with data from the COVID-19 Repository by the CSSE at Johns Hopkins
    \cite{johnshopkins_data, csse_jh_alsocite}. The first data point for each state shown correspond to the first date on which 1 or more cases were recorded. Raw data of cumulative case counts are used, and plotted on the same range of epidemic generations for purposes of comparison, despite an evident variability in the duration of generation length. Using a serial interval of 4 days, progressive generations are shown along the horizontal axis (generation two corresponds to 8 days, for example). On the vertical axis, the cumulative case counts for each state are plotted. We see how a state's proclivity to the epidemic taking off changes over the course of successive generations. Several states such as California, Massachusetts, and Washington had a lower probability of epidemic survival early on, then crossed the band into a higher likelihood over a short time span. Although the data used in this figure does not take into account factors such as missing count data, it serves as a visualization of how sharply the interplay of generation of epidemic and cumulative cases demarcate the probability of the epidemic continuing. }
    \label{fig:covid_heatmap}
\end{figure}

In Fig. \ref{fig:contour_prob} we show the probability of epidemic survival (that is, the probability of an epidemic continuing to grow) with a fixed generation $g=4$ and fixed cumulative cases $s=16$ over a range of $R_0$ and $k$ values, highlighting parameter estimates for COVID-19 \cite{riou_althaus_2020}. 
Despite the relatively low number of cases after several generations, clearly affected by the lack of testing resources at the time, the chances of the epidemic stochastically dying out were already close to a simple coin flip.
In Fig. \ref{fig:covid_heatmap}, we show the inverse problems: fixing disease parameters and varying temporal variables. We set $R_0=2.5$ and $k=0.1$, falling within the range of values for COVID-19, and track seven US states over time to observe where their disease progression state falls in the probability space of epidemic survival.

Guided by the results shown in Fig. \ref{fig:timing1}, we proceed knowing that our model predicts generations to emerge in linear increments of time. 
We use the serial interval of 4 days, taken from the window for COVID-19 \cite{du2020serial} to correspond with successive generational emergence. 
We observe that several states hovered around a low probability of epidemic survival at low early cases, but very quickly crossed to a much higher bracket where natural extinction of the disease spread is virtually impossible. The states of Washington and Massachusetts each took only two generations to cross from sub to supercritical epidemic survival probability, even derived from limited data and poor testing at the time.
The extraordinary leap in epidemic probability from just one generation to the next explain, in part, why it was so hard for public health systems to react and adapt to the spread of COVID-19.

%
%
%
%
%
\section{Discussion\label{sec:discussion}}
%

Temporal models of disease spread often fall in one of three categories. (i) Compartmental models that are deterministic in nature as they rely on ordinary differential equations, where uncertainty only stems from our imperfect knowledge of model parameters, rather than from the inherent randomness of disease transmission. (ii) Complicated agent-based models that lose the tractability of analytical models, which require significant amount of data to parametrize and do not produce explicit likelihood of outcomes. (iii) Time series analyses that can produce probabilistic forecasts. This last approach can produce useful predictions by ignoring transmission mechanisms or contact structure, but that perspective also precludes it from evaluating potential interventions that affect individual parameters or contact structure.

In this paper, we have shown that analysis of branching processes often used to only study the final state of epidemic models can actually combine the strengths of these different approaches by including stochasticity, contact heterogeneity and even individual characteristics \cite{newman2002spread, kenah2007second, allard2017asymmetric}. The reason this framework is usually used to solely predict the probability and final size of an epidemic is that the mathematical treatment involves integrating over contacts and therefore time \cite{hebert2020beyond}. However, we provided a first demonstration that the predictions made over generations by the branching process are actually very close approximation of continuous time epidemic dynamics on equivalent contact networks. This result alone justifies a large body of work and creates a foundation for analytical, probabilistic, epidemic forecasts based on PGFs.

Our probabilistic and temporal forecasts allowed us to uncover the diversity of epidemic courses, in the form of an unusually broad distribution of potential transmission trees over time. 
We have also shown that these flat distributions emerge on both homogeneous (e.g. Erd\H{o}s-R\'{e}nyi graphs) and heterogeneous (e.g. scale-free) contact networks. 
This phenomenon is therefore driven by the stochasticity of disease transmission rather than by the complexity of the contact structure.
This broad likelihood of early disease incidence justifies our use of a stochastic branching process, whereas deterministic models would typically track only the average or expected number of cases which is a poor description of flat distributions.

Our framework currently rests on a few assumptions. By building our framework on a configuration model, we ignore potentially important structural correlations. The PGF framework itself can be extended, data permitting, to include such correlations like degree-degree assortativity \cite{VaquezAlexei2003Rtdo, hebertdufresne2013percolation}, clustering \cite{NewmanMEJ2009Rgwc, AllardA2012Bpoa}, and more general structures \cite{KarrerBrian2010Rgca, AllardAntoine2015Gaea}. All of these generalizations of the PGF framework still rely, at some level, on a treelike approximation, but this approach has been shown to capture most important network features \cite{MelnikSergey2011Tueo}.

We also assume that there are a finite number of active generations at any given time and that the distribution of contacts and transmission probability do not change over time. This first assumption was tested in Fig.~\ref{fig:timing1} where we show that a simple network-based serial interval provides a reasonable approximation for time of emergence of epidemic generations in the continuous dynamics, illustrating both why and how we can align the generation-based branching process with the underlying temporal dynamics.

Our assumptions on the constant contact patterns and transmissibility provide a great road map for future work. In Eq. \eqref{eq:generation}, we formulate our PGFs on a per generation basis, which would allow us to change these patterns over time to model adaptive behavior or top-down interventions (e.g. lockdowns limiting contacts or masks reducing transmissibility). Certain network interventions have been shown to alter the dynamics of epidemic outcomes in interesting ways, such as contact tracing \cite{RiziAbbasK2021ESaD, kojaku2021effectiveness} or vaccination roll-outs \cite{BurgioGiulio2021Hits, HiraokaTakayuki2021HIaE}. Specifically, when interventions are targeted around key individuals (e.g. hubs \cite{PhysRevLett.86.3682}) or affect different subset of the population differently \cite{allard2017asymmetric}, one can see the emergence of smeared transitions when epidemics mostly spread in specific subgraphs with subcritical spillover in other populations \cite{PhysRevResearch.1.013009}.
Modeling interventions under a generational PGF framework would provide probabilistic forecasts not only of disease dynamics but also of the impact and timing of particular interventions.

Importantly, our results on the diversity of epidemic courses highlight how little information can actually be gathered from early incidence data. In Fig.~\ref{fig:time}, we see that the same disease in the same population can be roughly as likely to produce 40 or 400 cases after 10 epidemic generations. 

Finally, our results on epidemic survival show how quickly a situation can move from an uncertain outbreak to supercritical exponential growth. Due to both the randomness of disease spread and the imperfect COVID-19 testing protocols from early 2020, most states in the US moved from below 20\% survival probability of the epidemic to above 80\% in about two epidemic generations (2 weeks or less).

Altogether, our results stress the danger of justifying a lack of intervention with slow trends in early disease spread data.
Little can be learned about transmission mechanisms and dynamics from the first few epidemic generations.
The distribution of epidemic courses is mostly driven by the inherent randomness of transmission, and the window in which the dynamics settle into their subcritical or supercritical behavior tends to be unfortunately narrow, which leaves little room for fast adaptive responses.

Faced now with emergence of variants of COVID-19 around the world, the current situation is reminiscent of the scenario in the state of Washington during January of 2020 \textemdash sporadic clusters of cases with an unclear growth trajectory. We see from the data in Washington, as well as many other states and countries, how quickly cases explode and what that means for the likelihood of controlling the epidemic without external intervention efforts. Slow initial disease growth does not preclude a rapid increase shortly thereafter.

\section*{Acknowledgments}
A.J.A. and L.H.-D. acknowledge financial support from the National Institutes of Health 1P20 GM125498-01 Centers of Biomedical Research Excellence Award. M.C.B. is supported as a Fellow of the National Science Foundation under NRT Award No. DGE-1735316 and N.J.R. is supported by the University of Vermont. A.A. acknowledges financial support from the Sentinelle Nord initiative of the Canada First Research Excellence Fund and from the Natural Sciences and Engineering Research Council of Canada (Project No. 2019-05183).


\begin{thebibliography}{78}%
\makeatletter
\providecommand \@ifxundefined [1]{%
 \@ifx{#1\undefined}
}%
\providecommand \@ifnum [1]{%
 \ifnum #1\expandafter \@firstoftwo
 \else \expandafter \@secondoftwo
 \fi
}%
\providecommand \@ifx [1]{%
 \ifx #1\expandafter \@firstoftwo
 \else \expandafter \@secondoftwo
 \fi
}%
\providecommand \natexlab [1]{#1}%
\providecommand \enquote  [1]{``#1''}%
\providecommand \bibnamefont  [1]{#1}%
\providecommand \bibfnamefont [1]{#1}%
\providecommand \citenamefont [1]{#1}%
\providecommand \href@noop [0]{\@secondoftwo}%
\providecommand \href [0]{\begingroup \@sanitize@url \@href}%
\providecommand \@href[1]{\@@startlink{#1}\@@href}%
\providecommand \@@href[1]{\endgroup#1\@@endlink}%
\providecommand \@sanitize@url [0]{\catcode `\\12\catcode `\$12\catcode
  `\&12\catcode `\#12\catcode `\^12\catcode `\_12\catcode `\%12\relax}%
\providecommand \@@startlink[1]{}%
\providecommand \@@endlink[0]{}%
\providecommand \url  [0]{\begingroup\@sanitize@url \@url }%
\providecommand \@url [1]{\endgroup\@href {#1}{\urlprefix }}%
\providecommand \urlprefix  [0]{URL }%
\providecommand \Eprint [0]{\href }%
\providecommand \doibase [0]{https://doi.org/}%
\providecommand \selectlanguage [0]{\@gobble}%
\providecommand \bibinfo  [0]{\@secondoftwo}%
\providecommand \bibfield  [0]{\@secondoftwo}%
\providecommand \translation [1]{[#1]}%
\providecommand \BibitemOpen [0]{}%
\providecommand \bibitemStop [0]{}%
\providecommand \bibitemNoStop [0]{.\EOS\space}%
\providecommand \EOS [0]{\spacefactor3000\relax}%
\providecommand \BibitemShut  [1]{\csname bibitem#1\endcsname}%
\let\auto@bib@innerbib\@empty
\bibitem [{was(2020)}]{wash_firstcase_cdc}%
  \BibitemOpen
  \href@noop {} {\bibinfo {title} {{{F}irst Travel-related Case of 2019 Novel
  Coronavirus Detected in United States}}},\ \bibinfo {howpublished}
  {\url{https://www.cdc.gov/media/releases/2020/p0121-novel-coronavirus-travel-case.html}}
  (\bibinfo {year} {2020})\BibitemShut {NoStop}%
\bibitem [{\citenamefont {Oxley}\ and\ \citenamefont {Ryan}(ch
  7)}]{feb19_cases}%
  \BibitemOpen
  \bibfield  {author} {\bibinfo {author} {\bibfnamefont {D.}~\bibnamefont
  {Oxley}}\ and\ \bibinfo {author} {\bibfnamefont {J.}~\bibnamefont {Ryan}},\
  }\href {https://bit.ly/3hl9ncx} {\bibfield  {journal} {\bibinfo  {journal}
  {KUOW News and Inf.}\ } (\bibinfo {year} {2020, March 7})}\BibitemShut
  {NoStop}%
\bibitem [{\citenamefont {Sullivan}(ch 3)}]{feb26_deaths_covid}%
  \BibitemOpen
  \bibfield  {author} {\bibinfo {author} {\bibfnamefont {O.}~\bibnamefont
  {Sullivan}},\ }\href
  {https://www.kirklandreporter.com/news/coronavirus-death-toll-rises-to-eight-in-washington/}
  {\bibfield  {journal} {\bibinfo  {journal} {Kirkland Reporter}\ } (\bibinfo
  {year} {2020, March 3})}\BibitemShut {NoStop}%
\bibitem [{\citenamefont {Sundell}(uary)}]{school_closed_news}%
  \BibitemOpen
  \bibfield  {author} {\bibinfo {author} {\bibfnamefont {A.}~\bibnamefont
  {Sundell}},\ }\href
  {https://www.king5.com/article/news/health/coronavirus/washington-coronavirus-update/281-e73682dc-dad7-4b6e-b0ec-d2234ff9e2e0}
  {\bibfield  {journal} {\bibinfo  {journal} {KING-TV}\ } (\bibinfo {year}
  {2020, February})}\BibitemShut {NoStop}%
\bibitem [{\citenamefont {Kermack}\ and\ \citenamefont
  {McKendrick}(1927)}]{kermack1927contributionA}%
  \BibitemOpen
  \bibfield  {author} {\bibinfo {author} {\bibfnamefont {W.~O.}\ \bibnamefont
  {Kermack}}\ and\ \bibinfo {author} {\bibfnamefont {A.~G.}\ \bibnamefont
  {McKendrick}},\ }\href {https://doi.org/10.1098/rspa.1927.0118} {\bibfield
  {journal} {\bibinfo  {journal} {Proc. R. Soc. Lond., A}\ }\textbf {\bibinfo
  {volume} {115}},\ \bibinfo {pages} {700} (\bibinfo {year}
  {1927})}\BibitemShut {NoStop}%
\bibitem [{\citenamefont {Kermack}\ and\ \citenamefont
  {McKendrick}(1932)}]{kermack1927contributionB}%
  \BibitemOpen
  \bibfield  {author} {\bibinfo {author} {\bibfnamefont {W.~O.}\ \bibnamefont
  {Kermack}}\ and\ \bibinfo {author} {\bibfnamefont {A.~G.}\ \bibnamefont
  {McKendrick}},\ }\href {https://doi.org/10.1098/rspa.1927.0118} {\bibfield
  {journal} {\bibinfo  {journal} {Proc. R. Soc. Lond., A}\ }\textbf {\bibinfo
  {volume} {138}},\ \bibinfo {pages} {55} (\bibinfo {year} {1932})}\BibitemShut
  {NoStop}%
\bibitem [{\citenamefont {Kermack}\ and\ \citenamefont
  {McKendrick}(1933)}]{kermack1927contributionC}%
  \BibitemOpen
  \bibfield  {author} {\bibinfo {author} {\bibfnamefont {W.~O.}\ \bibnamefont
  {Kermack}}\ and\ \bibinfo {author} {\bibfnamefont {A.~G.}\ \bibnamefont
  {McKendrick}},\ }\href {https://doi.org/10.1098/rspa.1927.0118} {\bibfield
  {journal} {\bibinfo  {journal} {Proc. R. Soc. Lond., A}\ }\textbf {\bibinfo
  {volume} {141}},\ \bibinfo {pages} {94} (\bibinfo {year} {1933})}\BibitemShut
  {NoStop}%
\bibitem [{\citenamefont {Anderson}\ and\ \citenamefont
  {May}(1991)}]{AndersonMay}%
  \BibitemOpen
  \bibfield  {author} {\bibinfo {author} {\bibfnamefont {R.~M.}\ \bibnamefont
  {Anderson}}\ and\ \bibinfo {author} {\bibfnamefont {R.~M.}\ \bibnamefont
  {May}},\ }\href@noop {} {\emph {\bibinfo {title} {Infectious Diseases of
  Humans: Dynamics and Control}}}\ (\bibinfo  {publisher} {Oxf. Univ. Press},\
  \bibinfo {address} {Great Clarendon Street, Oxford OX2 6DP},\ \bibinfo {year}
  {1991})\BibitemShut {NoStop}%
\bibitem [{\citenamefont {Keeling}\ and\ \citenamefont
  {Rohani}(2007)}]{KeelingRohani}%
  \BibitemOpen
  \bibfield  {author} {\bibinfo {author} {\bibfnamefont {M.~J.}\ \bibnamefont
  {Keeling}}\ and\ \bibinfo {author} {\bibfnamefont {P.}~\bibnamefont
  {Rohani}},\ }\href {https://doi.org/10.2307/j.ctvcm4gk0} {\emph {\bibinfo
  {title} {Modeling Infectious Diseases in Humans and Animals}}}\ (\bibinfo
  {publisher} {Princet. Univ. Press},\ \bibinfo {address} {41 Williams St,
  Princeton, New Jersey 08540},\ \bibinfo {year} {2007})\BibitemShut {NoStop}%
\bibitem [{\citenamefont {Pastor-Satorras}\ \emph {et~al.}(2015)\citenamefont
  {Pastor-Satorras}, \citenamefont {Castellano}, \citenamefont {Van~Mieghem},\
  and\ \citenamefont {Vespignani}}]{pastor2015epidemic}%
  \BibitemOpen
  \bibfield  {author} {\bibinfo {author} {\bibfnamefont {R.}~\bibnamefont
  {Pastor-Satorras}}, \bibinfo {author} {\bibfnamefont {C.}~\bibnamefont
  {Castellano}}, \bibinfo {author} {\bibfnamefont {P.}~\bibnamefont
  {Van~Mieghem}},\ and\ \bibinfo {author} {\bibfnamefont {A.}~\bibnamefont
  {Vespignani}},\ }\href {https://doi.org/10.1103/RevModPhys.87.925} {\bibfield
   {journal} {\bibinfo  {journal} {Rev. Mod. Phys.}\ }\textbf {\bibinfo
  {volume} {87}},\ \bibinfo {pages} {925} (\bibinfo {year} {2015})}\BibitemShut
  {NoStop}%
\bibitem [{\citenamefont {H.}(1990)}]{Ibana1990}%
  \BibitemOpen
  \bibfield  {author} {\bibinfo {author} {\bibfnamefont {I.}~\bibnamefont
  {H.}},\ }\href {https://doi.org/https://doi.org/10.1007/BF00178326}
  {\bibfield  {journal} {\bibinfo  {journal} {J. Math. Biol.}\ }\textbf
  {\bibinfo {volume} {28}},\ \bibinfo {pages} {411} (\bibinfo {year}
  {1990})}\BibitemShut {NoStop}%
\bibitem [{\citenamefont {Huang}\ \emph {et~al.}(1992)\citenamefont {Huang},
  \citenamefont {Cooke},\ and\ \citenamefont {Castillo-Chavez}}]{Huang1992}%
  \BibitemOpen
  \bibfield  {author} {\bibinfo {author} {\bibfnamefont {W.}~\bibnamefont
  {Huang}}, \bibinfo {author} {\bibfnamefont {K.~L.}\ \bibnamefont {Cooke}},\
  and\ \bibinfo {author} {\bibfnamefont {C.}~\bibnamefont {Castillo-Chavez}},\
  }\href {https://doi.org/10.1137/0152047} {\bibfield  {journal} {\bibinfo
  {journal} {SIAM J. Appl. Math}\ }\textbf {\bibinfo {volume} {52}},\ \bibinfo
  {pages} {835} (\bibinfo {year} {1992})}\BibitemShut {NoStop}%
\bibitem [{\citenamefont {Bolker}\ and\ \citenamefont
  {Grenfell}(2015)}]{Bolker1995}%
  \BibitemOpen
  \bibfield  {author} {\bibinfo {author} {\bibfnamefont {B.}~\bibnamefont
  {Bolker}}\ and\ \bibinfo {author} {\bibfnamefont {B.}~\bibnamefont
  {Grenfell}},\ }\href {https://doi.org/10.1098/rstb.1995.0070} {\bibfield
  {journal} {\bibinfo  {journal} {Philos. Trans. R. Soc. B}\ }\textbf {\bibinfo
  {volume} {348}},\ \bibinfo {pages} {309} (\bibinfo {year}
  {2015})}\BibitemShut {NoStop}%
\bibitem [{\citenamefont {Lloyd}\ and\ \citenamefont
  {Jansen}(2004)}]{Lloyd2004}%
  \BibitemOpen
  \bibfield  {author} {\bibinfo {author} {\bibfnamefont {A.}~\bibnamefont
  {Lloyd}}\ and\ \bibinfo {author} {\bibfnamefont {V.}~\bibnamefont {Jansen}},\
  }\href {https://doi.org/https://doi.org/10.1016/j.mbs.2003.09.003} {\bibfield
   {journal} {\bibinfo  {journal} {Math. Biosci.}\ }\textbf {\bibinfo {volume}
  {188}},\ \bibinfo {pages} {1} (\bibinfo {year} {2004})}\BibitemShut {NoStop}%
\bibitem [{\citenamefont {Ball}\ and\ \citenamefont {P.}(1995)}]{Ball1995}%
  \BibitemOpen
  \bibfield  {author} {\bibinfo {author} {\bibfnamefont {F.}~\bibnamefont
  {Ball}}\ and\ \bibinfo {author} {\bibfnamefont {D.}~\bibnamefont {P.}},\
  }\href {https://doi.org/https://doi.org/10.1016/0304-4149(94)00034-Q}
  {\bibfield  {journal} {\bibinfo  {journal} {Stoch. Process. Their Appl.}\
  }\textbf {\bibinfo {volume} {55}},\ \bibinfo {pages} {1} (\bibinfo {year}
  {1995})}\BibitemShut {NoStop}%
\bibitem [{\citenamefont {Ball}\ \emph {et~al.}(1997)\citenamefont {Ball},
  \citenamefont {Mollison},\ and\ \citenamefont {Scalia-Tomba}}]{Ball1997}%
  \BibitemOpen
  \bibfield  {author} {\bibinfo {author} {\bibfnamefont {F.}~\bibnamefont
  {Ball}}, \bibinfo {author} {\bibfnamefont {D.}~\bibnamefont {Mollison}},\
  and\ \bibinfo {author} {\bibfnamefont {G.}~\bibnamefont {Scalia-Tomba}},\
  }\href {https://doi.org/10.1214/aoap/1034625252} {\bibfield  {journal}
  {\bibinfo  {journal} {Ann. Appl. Probab.}\ }\textbf {\bibinfo {volume} {7}},\
  \bibinfo {pages} {46} (\bibinfo {year} {1997})}\BibitemShut {NoStop}%
\bibitem [{\citenamefont {Allen}(2015)}]{L_Allen2015}%
  \BibitemOpen
  \bibfield  {author} {\bibinfo {author} {\bibfnamefont {L.}~\bibnamefont
  {Allen}},\ }\href {https://doi.org/https://doi.org/10.1007/978-3-319-21554-9}
  {\emph {\bibinfo {title} {Stochastic Population and Epidemic Models}}}\
  (\bibinfo  {publisher} {Springer},\ \bibinfo {year} {2015})\BibitemShut
  {NoStop}%
\bibitem [{\citenamefont {Allen}(2017)}]{LAllen2017}%
  \BibitemOpen
  \bibfield  {author} {\bibinfo {author} {\bibfnamefont {L.}~\bibnamefont
  {Allen}},\ }\href
  {https://doi.org/http://dx.doi.org/10.1016/j.idm.2017.03.001} {\bibfield
  {journal} {\bibinfo  {journal} {Infect. Dis. Model.}\ }\textbf {\bibinfo
  {volume} {2}},\ \bibinfo {pages} {128} (\bibinfo {year} {2017})}\BibitemShut
  {NoStop}%
\bibitem [{\citenamefont {Wang}\ \emph {et~al.}(2018)\citenamefont {Wang},
  \citenamefont {Cai}, \citenamefont {Ding},\ and\ \citenamefont
  {Gui}}]{Wang2018}%
  \BibitemOpen
  \bibfield  {author} {\bibinfo {author} {\bibfnamefont {W.}~\bibnamefont
  {Wang}}, \bibinfo {author} {\bibfnamefont {Y.}~\bibnamefont {Cai}}, \bibinfo
  {author} {\bibfnamefont {Z.}~\bibnamefont {Ding}},\ and\ \bibinfo {author}
  {\bibfnamefont {Z.}~\bibnamefont {Gui}},\ }\bibfield  {journal} {\bibinfo
  {journal} {Phys. A: Stat. Mech. Appl.}\ }\textbf {\bibinfo {volume} {509}},\
  \href {https://doi.org/http://dx.doi.org/10.1016/j.physa.2018.06.099}
  {http://dx.doi.org/10.1016/j.physa.2018.06.099} (\bibinfo {year}
  {2018})\BibitemShut {NoStop}%
\bibitem [{\citenamefont {Gray}\ \emph {et~al.}(2011)\citenamefont {Gray},
  \citenamefont {Greenhalgh}, \citenamefont {Hu}, \citenamefont {Mao},\ and\
  \citenamefont {Pan}}]{Gray2011}%
  \BibitemOpen
  \bibfield  {author} {\bibinfo {author} {\bibfnamefont {D.}~\bibnamefont
  {Gray}}, \bibinfo {author} {\bibfnamefont {L.}~\bibnamefont {Greenhalgh}},
  \bibinfo {author} {\bibfnamefont {L.}~\bibnamefont {Hu}}, \bibinfo {author}
  {\bibfnamefont {X.}~\bibnamefont {Mao}},\ and\ \bibinfo {author}
  {\bibfnamefont {J.}~\bibnamefont {Pan}},\ }\href
  {https://doi.org/https://doi.org/10.1137/10081856X} {\bibfield  {journal}
  {\bibinfo  {journal} {SIAM J. Appl. Math.}\ }\textbf {\bibinfo {volume}
  {71}},\ \bibinfo {pages} {876} (\bibinfo {year} {2011})}\BibitemShut
  {NoStop}%
\bibitem [{\citenamefont {Finkenst\"adt}\ and\ \citenamefont
  {Grenfell}(2000)}]{Finkenstadt2000}%
  \BibitemOpen
  \bibfield  {author} {\bibinfo {author} {\bibfnamefont {B.~F.}\ \bibnamefont
  {Finkenst\"adt}}\ and\ \bibinfo {author} {\bibfnamefont {B.}~\bibnamefont
  {Grenfell}},\ }\href {https://doi.org/10.1111/1467-9876.00187} {\bibfield
  {journal} {\bibinfo  {journal} {Appl. Stat.}\ }\textbf {\bibinfo {volume}
  {49}},\ \bibinfo {pages} {187} (\bibinfo {year} {2000})}\BibitemShut
  {NoStop}%
\bibitem [{\citenamefont {Zhan}\ \emph {et~al.}(2018)\citenamefont {Zhan},
  \citenamefont {Dong}, \citenamefont {Lu}, \citenamefont {Yang}, \citenamefont
  {Wang},\ and\ \citenamefont {Jia}}]{Zhan2018}%
  \BibitemOpen
  \bibfield  {author} {\bibinfo {author} {\bibfnamefont {Z.}~\bibnamefont
  {Zhan}}, \bibinfo {author} {\bibfnamefont {W.}~\bibnamefont {Dong}}, \bibinfo
  {author} {\bibfnamefont {Y.}~\bibnamefont {Lu}}, \bibinfo {author}
  {\bibfnamefont {P.}~\bibnamefont {Yang}}, \bibinfo {author} {\bibfnamefont
  {Q.}~\bibnamefont {Wang}},\ and\ \bibinfo {author} {\bibfnamefont
  {P.}~\bibnamefont {Jia}},\ }\href
  {https://doi.org/10.1038/s41598-019-38930-y} {\bibfield  {journal} {\bibinfo
  {journal} {Sci. Rep.}\ }\textbf {\bibinfo {volume} {9}},\ \bibinfo {pages}
  {1} (\bibinfo {year} {2018})}\BibitemShut {NoStop}%
\bibitem [{\citenamefont {Allard}(1998)}]{Allard1998}%
  \BibitemOpen
  \bibfield  {author} {\bibinfo {author} {\bibfnamefont {R.}~\bibnamefont
  {Allard}},\ }\href@noop {} {\bibfield  {journal} {\bibinfo  {journal} {Bull.
  World Health Organ.}\ }\textbf {\bibinfo {volume} {76}},\ \bibinfo {pages}
  {327} (\bibinfo {year} {1998})}\BibitemShut {NoStop}%
\bibitem [{\citenamefont {Lopman}\ \emph {et~al.}(2009)\citenamefont {Lopman},
  \citenamefont {Armstrong}, \citenamefont {Atchinson},\ and\ \citenamefont
  {Gray}}]{Lopman2009}%
  \BibitemOpen
  \bibfield  {author} {\bibinfo {author} {\bibfnamefont {B.}~\bibnamefont
  {Lopman}}, \bibinfo {author} {\bibfnamefont {B.}~\bibnamefont {Armstrong}},
  \bibinfo {author} {\bibfnamefont {C.}~\bibnamefont {Atchinson}},\ and\
  \bibinfo {author} {\bibfnamefont {J.~J.}\ \bibnamefont {Gray}},\ }\href
  {https://doi.org/10.1371/journal.pone.0006671} {\bibfield  {journal}
  {\bibinfo  {journal} {PLoS One}\ }\textbf {\bibinfo {volume} {4}},\ \bibinfo
  {pages} {e6671} (\bibinfo {year} {2009})}\BibitemShut {NoStop}%
\bibitem [{\citenamefont {Hu}\ \emph {et~al.}(2006)\citenamefont {Hu},
  \citenamefont {Tong}, \citenamefont {Mengersen},\ and\ \citenamefont
  {Oldenburg}}]{Hu2006}%
  \BibitemOpen
  \bibfield  {author} {\bibinfo {author} {\bibfnamefont {W.}~\bibnamefont
  {Hu}}, \bibinfo {author} {\bibfnamefont {S.}~\bibnamefont {Tong}}, \bibinfo
  {author} {\bibfnamefont {K.}~\bibnamefont {Mengersen}},\ and\ \bibinfo
  {author} {\bibfnamefont {B.}~\bibnamefont {Oldenburg}},\ }\href
  {https://doi.org/10.1016/j.ecolmodel.2006.02.028} {\bibfield  {journal}
  {\bibinfo  {journal} {Ecol. Model.}\ }\textbf {\bibinfo {volume} {196}},\
  \bibinfo {pages} {505} (\bibinfo {year} {2006})}\BibitemShut {NoStop}%
\bibitem [{\citenamefont {Silva}\ \emph {et~al.}(2020)\citenamefont {Silva},
  \citenamefont {Batista}, \citenamefont {Lima}, \citenamefont {Alves},
  \citenamefont {Guimar{\~a}es},\ and\ \citenamefont {Silva}}]{silva2020covid}%
  \BibitemOpen
  \bibfield  {author} {\bibinfo {author} {\bibfnamefont {P.~C.~L.}\
  \bibnamefont {Silva}}, \bibinfo {author} {\bibfnamefont {P.~V.~C.}\
  \bibnamefont {Batista}}, \bibinfo {author} {\bibfnamefont {H.~S.}\
  \bibnamefont {Lima}}, \bibinfo {author} {\bibfnamefont {M.~A.}\ \bibnamefont
  {Alves}}, \bibinfo {author} {\bibfnamefont {F.~G.}\ \bibnamefont
  {Guimar{\~a}es}},\ and\ \bibinfo {author} {\bibfnamefont {R.~C.~P.}\
  \bibnamefont {Silva}},\ }\href {https://doi.org/10.1016/j.chaos.2020.110088}
  {\bibfield  {journal} {\bibinfo  {journal} {Chaos Solitons Fractals}\
  }\textbf {\bibinfo {volume} {139}},\ \bibinfo {pages} {110088} (\bibinfo
  {year} {2020})}\BibitemShut {NoStop}%
\bibitem [{\citenamefont {Gharakhanlou}\ and\ \citenamefont
  {Hooshangi}(2020)}]{gharakhanlou2020spatio}%
  \BibitemOpen
  \bibfield  {author} {\bibinfo {author} {\bibfnamefont {N.~M.}\ \bibnamefont
  {Gharakhanlou}}\ and\ \bibinfo {author} {\bibfnamefont {N.}~\bibnamefont
  {Hooshangi}},\ }\href {https://doi.org/10.1016/j.imu.2020.100403} {\bibfield
  {journal} {\bibinfo  {journal} {Inform. Med. Unlocked}\ }\textbf {\bibinfo
  {volume} {20}},\ \bibinfo {pages} {100403} (\bibinfo {year}
  {2020})}\BibitemShut {NoStop}%
\bibitem [{\citenamefont {Cuevas}(2020)}]{cuevas2020agent}%
  \BibitemOpen
  \bibfield  {author} {\bibinfo {author} {\bibfnamefont {E.}~\bibnamefont
  {Cuevas}},\ }\href {https://doi.org/10.1016/j.compbiomed.2020.103827}
  {\bibfield  {journal} {\bibinfo  {journal} {Comput. biol. med.}\ }\textbf
  {\bibinfo {volume} {121}},\ \bibinfo {pages} {103827} (\bibinfo {year}
  {2020})}\BibitemShut {NoStop}%
\bibitem [{\citenamefont {Hoertel}\ \emph {et~al.}(2020)\citenamefont
  {Hoertel}, \citenamefont {Blachier}, \citenamefont {Blanco}, \citenamefont
  {Olfson}, \citenamefont {Massetti}, \citenamefont {Limosin},\ and\
  \citenamefont {Leleu}}]{hoertel2020facing}%
  \BibitemOpen
  \bibfield  {author} {\bibinfo {author} {\bibfnamefont {N.}~\bibnamefont
  {Hoertel}}, \bibinfo {author} {\bibfnamefont {M.}~\bibnamefont {Blachier}},
  \bibinfo {author} {\bibfnamefont {C.}~\bibnamefont {Blanco}}, \bibinfo
  {author} {\bibfnamefont {M.}~\bibnamefont {Olfson}}, \bibinfo {author}
  {\bibfnamefont {M.}~\bibnamefont {Massetti}}, \bibinfo {author}
  {\bibfnamefont {F.}~\bibnamefont {Limosin}},\ and\ \bibinfo {author}
  {\bibfnamefont {H.}~\bibnamefont {Leleu}},\ }\href
  {https://doi.org/10.1101/2020.04.23.20076885} {\bibfield  {journal} {\bibinfo
   {journal} {MedRxiv}\ ,\ \bibinfo {pages} {1}} (\bibinfo {year}
  {2020})}\BibitemShut {NoStop}%
\bibitem [{\citenamefont {Staffini}\ \emph {et~al.}(2021)\citenamefont
  {Staffini}, \citenamefont {Svensson}, \citenamefont {Chung},\ and\
  \citenamefont {Svensson}}]{staffini2021agent}%
  \BibitemOpen
  \bibfield  {author} {\bibinfo {author} {\bibfnamefont {A.}~\bibnamefont
  {Staffini}}, \bibinfo {author} {\bibfnamefont {A.~K.}\ \bibnamefont
  {Svensson}}, \bibinfo {author} {\bibfnamefont {U.-I.}\ \bibnamefont
  {Chung}},\ and\ \bibinfo {author} {\bibfnamefont {T.}~\bibnamefont
  {Svensson}},\ }\href {https://doi.org/10.2196/24192} {\bibfield  {journal}
  {\bibinfo  {journal} {JMIR med. inform.}\ }\textbf {\bibinfo {volume} {9}},\
  \bibinfo {pages} {e24192} (\bibinfo {year} {2021})}\BibitemShut {NoStop}%
\bibitem [{\citenamefont {Srikrishnan}\ and\ \citenamefont
  {Keller}(2021)}]{srikrishnan2021small}%
  \BibitemOpen
  \bibfield  {author} {\bibinfo {author} {\bibfnamefont {V.}~\bibnamefont
  {Srikrishnan}}\ and\ \bibinfo {author} {\bibfnamefont {K.}~\bibnamefont
  {Keller}},\ }\href {https://doi.org/10.1016/j.envsoft.2021.104978} {\bibfield
   {journal} {\bibinfo  {journal} {Environ. Model. Softw.}\ }\textbf {\bibinfo
  {volume} {138}},\ \bibinfo {pages} {104978} (\bibinfo {year}
  {2021})}\BibitemShut {NoStop}%
\bibitem [{\citenamefont {Meyers}(2007)}]{meyers2007contact}%
  \BibitemOpen
  \bibfield  {author} {\bibinfo {author} {\bibfnamefont {L.}~\bibnamefont
  {Meyers}},\ }\href {https://doi.org/10.1090/S0273-0979-06-01148-7} {\bibfield
   {journal} {\bibinfo  {journal} {Bull New Ser. Am Math Soc.}\ }\textbf
  {\bibinfo {volume} {44}},\ \bibinfo {pages} {63} (\bibinfo {year}
  {2007})}\BibitemShut {NoStop}%
\bibitem [{\citenamefont {Kenah}\ and\ \citenamefont
  {Robins}(2007)}]{kenah2007second}%
  \BibitemOpen
  \bibfield  {author} {\bibinfo {author} {\bibfnamefont {E.}~\bibnamefont
  {Kenah}}\ and\ \bibinfo {author} {\bibfnamefont {J.~M.}\ \bibnamefont
  {Robins}},\ }\href {https://doi.org/10.1103/PhysRevE.76.036113} {\bibfield
  {journal} {\bibinfo  {journal} {Phys. Rev. E}\ }\textbf {\bibinfo {volume}
  {76}},\ \bibinfo {pages} {036113} (\bibinfo {year} {2007})}\BibitemShut
  {NoStop}%
\bibitem [{\citenamefont {Miller}(2007)}]{miller2007epidemic}%
  \BibitemOpen
  \bibfield  {author} {\bibinfo {author} {\bibfnamefont {J.~C.}\ \bibnamefont
  {Miller}},\ }\href {https://doi.org/10.1103/PhysRevE.76.010101} {\bibfield
  {journal} {\bibinfo  {journal} {Phys. Rev. E}\ }\textbf {\bibinfo {volume}
  {76}},\ \bibinfo {pages} {010101} (\bibinfo {year} {2007})}\BibitemShut
  {NoStop}%
\bibitem [{\citenamefont {Kenah}\ and\ \citenamefont
  {Miller}(2011)}]{kenah2011epidemic}%
  \BibitemOpen
  \bibfield  {author} {\bibinfo {author} {\bibfnamefont {E.}~\bibnamefont
  {Kenah}}\ and\ \bibinfo {author} {\bibfnamefont {J.~C.}\ \bibnamefont
  {Miller}},\ }\href {https://doi.org/10.1155/2011/543520} {\bibfield
  {journal} {\bibinfo  {journal} {Interdiscip. perspect. infect. dis.}\
  }\textbf {\bibinfo {volume} {2011}},\ \bibinfo {pages} {1} (\bibinfo {year}
  {2011})}\BibitemShut {NoStop}%
\bibitem [{\citenamefont {Athreya}\ and\ \citenamefont
  {Ney}(1972)}]{AthreyaNey}%
  \BibitemOpen
  \bibfield  {author} {\bibinfo {author} {\bibfnamefont {K.~B.}\ \bibnamefont
  {Athreya}}\ and\ \bibinfo {author} {\bibfnamefont {P.~E.}\ \bibnamefont
  {Ney}},\ }\href {https://doi.org/10.1007/978-3-642-65371-1} {\emph {\bibinfo
  {title} {Branching Processes}}}\ (\bibinfo  {publisher} {Springer-Verlag
  Berlin Heidelberg},\ \bibinfo {address} {New York},\ \bibinfo {year}
  {1972})\BibitemShut {NoStop}%
\bibitem [{\citenamefont {Newman}\ \emph {et~al.}(2001)\citenamefont {Newman},
  \citenamefont {Strogatz},\ and\ \citenamefont {Watts}}]{newman2001random}%
  \BibitemOpen
  \bibfield  {author} {\bibinfo {author} {\bibfnamefont {M.~E.~J.}\
  \bibnamefont {Newman}}, \bibinfo {author} {\bibfnamefont {S.~H.}\
  \bibnamefont {Strogatz}},\ and\ \bibinfo {author} {\bibfnamefont {D.~J.}\
  \bibnamefont {Watts}},\ }\href {https://doi.org/10.1103/PhysRevE.64.026118}
  {\bibfield  {journal} {\bibinfo  {journal} {Phys. Rev. E}\ }\textbf {\bibinfo
  {volume} {64}},\ \bibinfo {pages} {026118} (\bibinfo {year}
  {2001})}\BibitemShut {NoStop}%
\bibitem [{\citenamefont {Miller}(2018)}]{Miller2018bp}%
  \BibitemOpen
  \bibfield  {author} {\bibinfo {author} {\bibfnamefont {J.~C.}\ \bibnamefont
  {Miller}},\ }\href {https://doi.org/10.1016/j.idm.2018.08.001} {\bibfield
  {journal} {\bibinfo  {journal} {Infect. Dis. Model.}\ }\textbf {\bibinfo
  {volume} {3}},\ \bibinfo {pages} {192} (\bibinfo {year} {2018})}\BibitemShut
  {NoStop}%
\bibitem [{\citenamefont {Newman}(2002)}]{newman2002spread}%
  \BibitemOpen
  \bibfield  {author} {\bibinfo {author} {\bibfnamefont {M.~E.~J.}\
  \bibnamefont {Newman}},\ }\href {https://doi.org/10.1103/PhysRevE.66.016128}
  {\bibfield  {journal} {\bibinfo  {journal} {Phys. Rev. E}\ }\textbf {\bibinfo
  {volume} {66}},\ \bibinfo {pages} {016128} (\bibinfo {year}
  {2002})}\BibitemShut {NoStop}%
\bibitem [{\citenamefont {Levesque}\ \emph {et~al.}(2021)\citenamefont
  {Levesque}, \citenamefont {Maybury},\ and\ \citenamefont
  {Shaw}}]{levesque2021model}%
  \BibitemOpen
  \bibfield  {author} {\bibinfo {author} {\bibfnamefont {J.}~\bibnamefont
  {Levesque}}, \bibinfo {author} {\bibfnamefont {D.~W.}\ \bibnamefont
  {Maybury}},\ and\ \bibinfo {author} {\bibfnamefont {R.~D.}\ \bibnamefont
  {Shaw}},\ }\href {https://doi.org/10.1016/j.jtbi.2020.110536} {\bibfield
  {journal} {\bibinfo  {journal} {J. Theor. Biol.}\ }\textbf {\bibinfo {volume}
  {512}},\ \bibinfo {pages} {110536} (\bibinfo {year} {2021})}\BibitemShut
  {NoStop}%
\bibitem [{\citenamefont {Bertozzi}\ \emph {et~al.}(2020)\citenamefont
  {Bertozzi}, \citenamefont {Franco}, \citenamefont {Mohler}, \citenamefont
  {Short},\ and\ \citenamefont {Sledge}}]{bertozzi2020challenges}%
  \BibitemOpen
  \bibfield  {author} {\bibinfo {author} {\bibfnamefont {A.~L.}\ \bibnamefont
  {Bertozzi}}, \bibinfo {author} {\bibfnamefont {E.}~\bibnamefont {Franco}},
  \bibinfo {author} {\bibfnamefont {G.}~\bibnamefont {Mohler}}, \bibinfo
  {author} {\bibfnamefont {M.~B.}\ \bibnamefont {Short}},\ and\ \bibinfo
  {author} {\bibfnamefont {D.}~\bibnamefont {Sledge}},\ }\href
  {https://doi.org/10.1073/pnas.2006520117} {\bibfield  {journal} {\bibinfo
  {journal} {Proc. Natl. Acad. Sci.}\ }\textbf {\bibinfo {volume} {117}},\
  \bibinfo {pages} {16732} (\bibinfo {year} {2020})}\BibitemShut {NoStop}%
\bibitem [{\citenamefont {Mitrofani}\ and\ \citenamefont
  {Koutras}(2021)}]{mitrofani2021branching}%
  \BibitemOpen
  \bibfield  {author} {\bibinfo {author} {\bibfnamefont {I.~A.}\ \bibnamefont
  {Mitrofani}}\ and\ \bibinfo {author} {\bibfnamefont {V.~P.}\ \bibnamefont
  {Koutras}},\ }\bibfield  {journal} {\bibinfo  {journal} {Int. J. Model.
  Optim.}\ }\textbf {\bibinfo {volume} {11}},\ \href
  {https://doi.org/10.7763/IJMO.2021.V11.779} {10.7763/IJMO.2021.V11.779}
  (\bibinfo {year} {2021})\BibitemShut {NoStop}%
\bibitem [{\citenamefont {Zhang}\ \emph {et~al.}(2021)\citenamefont {Zhang},
  \citenamefont {Wang}, \citenamefont {Liu}, \citenamefont {Liu}, \citenamefont
  {Feng},\ and\ \citenamefont {Wu}}]{zhang2021heterogeneous}%
  \BibitemOpen
  \bibfield  {author} {\bibinfo {author} {\bibfnamefont {L.}~\bibnamefont
  {Zhang}}, \bibinfo {author} {\bibfnamefont {H.}~\bibnamefont {Wang}},
  \bibinfo {author} {\bibfnamefont {Z.}~\bibnamefont {Liu}}, \bibinfo {author}
  {\bibfnamefont {X.~F.}\ \bibnamefont {Liu}}, \bibinfo {author} {\bibfnamefont
  {X.}~\bibnamefont {Feng}},\ and\ \bibinfo {author} {\bibfnamefont
  {Y.}~\bibnamefont {Wu}},\ }\href {https://doi.org/10.1155/2021/6686547}
  {\bibfield  {journal} {\bibinfo  {journal} {Complexity}\ }\textbf {\bibinfo
  {volume} {2021}},\ \bibinfo {pages} {1} (\bibinfo {year} {2021})}\BibitemShut
  {NoStop}%
\bibitem [{\citenamefont {Akian}\ \emph {et~al.}(2020)\citenamefont {Akian},
  \citenamefont {Ganassali}, \citenamefont {Gaubert},\ and\ \citenamefont
  {Massouli{\'e}}}]{akian2020probabilistic}%
  \BibitemOpen
  \bibfield  {author} {\bibinfo {author} {\bibfnamefont {M.}~\bibnamefont
  {Akian}}, \bibinfo {author} {\bibfnamefont {L.}~\bibnamefont {Ganassali}},
  \bibinfo {author} {\bibfnamefont {S.}~\bibnamefont {Gaubert}},\ and\ \bibinfo
  {author} {\bibfnamefont {L.}~\bibnamefont {Massouli{\'e}}},\ }\href
  {https://doi.org/arXiv:2009.05304} {\bibfield  {journal} {\bibinfo  {journal}
  {arXiv}\ ,\ \bibinfo {pages} {1}} (\bibinfo {year} {2020})}\BibitemShut
  {NoStop}%
\bibitem [{\citenamefont {Kojaku}\ \emph {et~al.}(2021)\citenamefont {Kojaku},
  \citenamefont {H{\'e}bert-Dufresne}, \citenamefont {Mones}, \citenamefont
  {Lehmann},\ and\ \citenamefont {Ahn}}]{kojaku2021effectiveness}%
  \BibitemOpen
  \bibfield  {author} {\bibinfo {author} {\bibfnamefont {S.}~\bibnamefont
  {Kojaku}}, \bibinfo {author} {\bibfnamefont {L.}~\bibnamefont
  {H{\'e}bert-Dufresne}}, \bibinfo {author} {\bibfnamefont {E.}~\bibnamefont
  {Mones}}, \bibinfo {author} {\bibfnamefont {S.}~\bibnamefont {Lehmann}},\
  and\ \bibinfo {author} {\bibfnamefont {Y.-Y.}\ \bibnamefont {Ahn}},\
  }\href@noop {} {\bibfield  {journal} {\bibinfo  {journal} {Nature Physics}\
  }\textbf {\bibinfo {volume} {17}},\ \bibinfo {pages} {652} (\bibinfo {year}
  {2021})}\BibitemShut {NoStop}%
\bibitem [{\citenamefont {No{\"e}l}\ \emph {et~al.}(2009)\citenamefont
  {No{\"e}l}, \citenamefont {Davoudi}, \citenamefont {Brunham}, \citenamefont
  {Dub{\'e}},\ and\ \citenamefont {Pourbohloul}}]{noel2009time}%
  \BibitemOpen
  \bibfield  {author} {\bibinfo {author} {\bibfnamefont {P.-A.}\ \bibnamefont
  {No{\"e}l}}, \bibinfo {author} {\bibfnamefont {B.}~\bibnamefont {Davoudi}},
  \bibinfo {author} {\bibfnamefont {R.~C.}\ \bibnamefont {Brunham}}, \bibinfo
  {author} {\bibfnamefont {L.~J.}\ \bibnamefont {Dub{\'e}}},\ and\ \bibinfo
  {author} {\bibfnamefont {B.}~\bibnamefont {Pourbohloul}},\ }\href
  {https://doi.org/10.1103/PhysRevE.79.026101} {\bibfield  {journal} {\bibinfo
  {journal} {Phys. Rev. E}\ }\textbf {\bibinfo {volume} {79}},\ \bibinfo
  {pages} {026101} (\bibinfo {year} {2009})}\BibitemShut {NoStop}%
\bibitem [{\citenamefont {Wilf}(2005)}]{wilf2005generatingfunctionology}%
  \BibitemOpen
  \bibfield  {author} {\bibinfo {author} {\bibfnamefont {H.~S.}\ \bibnamefont
  {Wilf}},\ }\href@noop {} {\emph {\bibinfo {title}
  {generatingfunctionology}}}\ (\bibinfo  {publisher} {CRC press},\ \bibinfo
  {address} {Boca Raton, Florida},\ \bibinfo {year} {2005})\BibitemShut
  {NoStop}%
\bibitem [{\citenamefont {Fosdick}\ \emph {et~al.}(2018)\citenamefont
  {Fosdick}, \citenamefont {Larremore}, \citenamefont {Nishimura},\ and\
  \citenamefont {Ugander}}]{FosdickBaileyK2018CRGM}%
  \BibitemOpen
  \bibfield  {author} {\bibinfo {author} {\bibfnamefont {B.~K.}\ \bibnamefont
  {Fosdick}}, \bibinfo {author} {\bibfnamefont {D.~B.}\ \bibnamefont
  {Larremore}}, \bibinfo {author} {\bibfnamefont {J.}~\bibnamefont
  {Nishimura}},\ and\ \bibinfo {author} {\bibfnamefont {J.}~\bibnamefont
  {Ugander}},\ }\href@noop {} {\bibfield  {journal} {\bibinfo  {journal} {SIAM
  review}\ }\textbf {\bibinfo {volume} {60}},\ \bibinfo {pages} {315} (\bibinfo
  {year} {2018})}\BibitemShut {NoStop}%
\bibitem [{and(2021)}]{andrea_allen_2021_5076514}%
  \BibitemOpen
  \href {https://doi.org/10.5281/zenodo.5076514} {\bibinfo {title}
  {{andrea-allen/epintervene: Initial Release v1.0.0}}} (\bibinfo {year}
  {2021}),\ \bibinfo {note} {10.5281/zenodo.5076514}\BibitemShut {NoStop}%
\bibitem [{\citenamefont {Kiss}\ \emph {et~al.}(2019)\citenamefont {Kiss},
  \citenamefont {Miller},\ and\ \citenamefont
  {Simon}}]{kissmathonnetworks2019}%
  \BibitemOpen
  \bibfield  {author} {\bibinfo {author} {\bibfnamefont {I.~Z.}\ \bibnamefont
  {Kiss}}, \bibinfo {author} {\bibfnamefont {J.~C.}\ \bibnamefont {Miller}},\
  and\ \bibinfo {author} {\bibfnamefont {P.~L.}\ \bibnamefont {Simon}},\ }\href
  {https://doi.org/10.1007/978-3-319-50806-1} {\emph {\bibinfo {title}
  {Mathematics of Epidemics on Networks: from exact to approximate models}}}\
  (\bibinfo  {publisher} {Springer},\ \bibinfo {year} {2019})\BibitemShut
  {NoStop}%
\bibitem [{\citenamefont {Miller}\ and\ \citenamefont
  {Ting}(2019)}]{miller_ting_2019}%
  \BibitemOpen
  \bibfield  {author} {\bibinfo {author} {\bibfnamefont {J.~C.}\ \bibnamefont
  {Miller}}\ and\ \bibinfo {author} {\bibfnamefont {T.}~\bibnamefont {Ting}},\
  }\href {https://doi.org/10.21105/joss.01731} {\bibfield  {journal} {\bibinfo
  {journal} {J. Open Source Softw.}\ }\textbf {\bibinfo {volume} {4}},\
  \bibinfo {pages} {1731} (\bibinfo {year} {2019})}\BibitemShut {NoStop}%
\bibitem [{\citenamefont {Bauer}\ \emph {et~al.}(2016)\citenamefont {Bauer},
  \citenamefont {Engblom},\ and\ \citenamefont
  {Widgren}}]{bauer_engblom_widgren_2016}%
  \BibitemOpen
  \bibfield  {author} {\bibinfo {author} {\bibfnamefont {P.}~\bibnamefont
  {Bauer}}, \bibinfo {author} {\bibfnamefont {S.}~\bibnamefont {Engblom}},\
  and\ \bibinfo {author} {\bibfnamefont {S.}~\bibnamefont {Widgren}},\ }\href
  {https://doi.org/10.1177/1094342016635723} {\bibfield  {journal} {\bibinfo
  {journal} {Inter, J. High Perform. Comput. Appl.}\ }\textbf {\bibinfo
  {volume} {30}},\ \bibinfo {pages} {438} (\bibinfo {year} {2016})}\BibitemShut
  {NoStop}%
\bibitem [{\citenamefont {Gillespie}(1977)}]{gillespie1977}%
  \BibitemOpen
  \bibfield  {author} {\bibinfo {author} {\bibfnamefont {D.~T.}\ \bibnamefont
  {Gillespie}},\ }\href {https://doi.org/10.1021/j100540a008} {\bibfield
  {journal} {\bibinfo  {journal} {J. Phys. Chem.}\ }\textbf {\bibinfo {volume}
  {81}},\ \bibinfo {pages} {2340} (\bibinfo {year} {1977})}\BibitemShut
  {NoStop}%
\bibitem [{\citenamefont {H{\'e}bert-Dufresne}\ \emph
  {et~al.}(2020)\citenamefont {H{\'e}bert-Dufresne}, \citenamefont {Althouse},
  \citenamefont {Scarpino},\ and\ \citenamefont {Allard}}]{hebert2020beyond}%
  \BibitemOpen
  \bibfield  {author} {\bibinfo {author} {\bibfnamefont {L.}~\bibnamefont
  {H{\'e}bert-Dufresne}}, \bibinfo {author} {\bibfnamefont {B.~M.}\
  \bibnamefont {Althouse}}, \bibinfo {author} {\bibfnamefont {S.~V.}\
  \bibnamefont {Scarpino}},\ and\ \bibinfo {author} {\bibfnamefont
  {A.}~\bibnamefont {Allard}},\ }\href {https://doi.org/10.1098/rsif.2020.0393}
  {\bibfield  {journal} {\bibinfo  {journal} {J. R. Soc. Interface}\ }\textbf
  {\bibinfo {volume} {17}},\ \bibinfo {pages} {20200393} (\bibinfo {year}
  {2020})}\BibitemShut {NoStop}%
\bibitem [{\citenamefont {Du}\ \emph {et~al.}(2020)\citenamefont {Du},
  \citenamefont {Xu}, \citenamefont {Wu}, \citenamefont {Wang}, \citenamefont
  {Cowling},\ and\ \citenamefont {Meyers}}]{du2020serial}%
  \BibitemOpen
  \bibfield  {author} {\bibinfo {author} {\bibfnamefont {Z.}~\bibnamefont
  {Du}}, \bibinfo {author} {\bibfnamefont {X.}~\bibnamefont {Xu}}, \bibinfo
  {author} {\bibfnamefont {Y.}~\bibnamefont {Wu}}, \bibinfo {author}
  {\bibfnamefont {L.}~\bibnamefont {Wang}}, \bibinfo {author} {\bibfnamefont
  {B.~J.}\ \bibnamefont {Cowling}},\ and\ \bibinfo {author} {\bibfnamefont
  {L.~A.}\ \bibnamefont {Meyers}},\ }\href
  {https://doi.org/10.3201/eid2606.200357} {\bibfield  {journal} {\bibinfo
  {journal} {Emerg Infect Dis.}\ }\textbf {\bibinfo {volume} {26}},\ \bibinfo
  {pages} {1341} (\bibinfo {year} {2020})}\BibitemShut {NoStop}%
\bibitem [{\citenamefont {Becker}(1977)}]{Becker77}%
  \BibitemOpen
  \bibfield  {author} {\bibinfo {author} {\bibfnamefont {N.~G.}\ \bibnamefont
  {Becker}},\ }\href {https://doi.org/10.2307/2529366} {\bibfield  {journal}
  {\bibinfo  {journal} {Biometrics}\ }\textbf {\bibinfo {volume} {33}},\
  \bibinfo {pages} {515} (\bibinfo {year} {1977})}\BibitemShut {NoStop}%
\bibitem [{\citenamefont {Becker}(1974)}]{Becker74}%
  \BibitemOpen
  \bibfield  {author} {\bibinfo {author} {\bibfnamefont {N.~G.}\ \bibnamefont
  {Becker}},\ }\href {https://doi.org/10.2307/2334720} {\bibfield  {journal}
  {\bibinfo  {journal} {Biometrika}\ }\textbf {\bibinfo {volume} {61}},\
  \bibinfo {pages} {393} (\bibinfo {year} {1974})}\BibitemShut {NoStop}%
\bibitem [{\citenamefont {Diekmann}\ \emph {et~al.}(2013)\citenamefont
  {Diekmann}, \citenamefont {Heesterbeek},\ and\ \citenamefont
  {Britton}}]{DHB_book}%
  \BibitemOpen
  \bibfield  {author} {\bibinfo {author} {\bibfnamefont {O.}~\bibnamefont
  {Diekmann}}, \bibinfo {author} {\bibfnamefont {H.}~\bibnamefont
  {Heesterbeek}},\ and\ \bibinfo {author} {\bibfnamefont {T.}~\bibnamefont
  {Britton}},\ }\href@noop {} {\emph {\bibinfo {title} {Mathematical Tools for
  Understanding Infectious Disease Dynamics}}}\ (\bibinfo  {publisher}
  {Princet. Univ. Press},\ \bibinfo {address} {41 Williams St, Princeton, New
  Jersey 08540},\ \bibinfo {year} {2013})\BibitemShut {NoStop}%
\bibitem [{\citenamefont {Lloyd-Smith}\ \emph {et~al.}(2005)\citenamefont
  {Lloyd-Smith}, \citenamefont {Schreiber}, \citenamefont {Kopp},\ and\
  \citenamefont {Getz}}]{lloyd-smith_schreiber_kopp_getz_2005}%
  \BibitemOpen
  \bibfield  {author} {\bibinfo {author} {\bibfnamefont {J.~O.}\ \bibnamefont
  {Lloyd-Smith}}, \bibinfo {author} {\bibfnamefont {S.~J.}\ \bibnamefont
  {Schreiber}}, \bibinfo {author} {\bibfnamefont {P.~E.}\ \bibnamefont
  {Kopp}},\ and\ \bibinfo {author} {\bibfnamefont {W.~M.}\ \bibnamefont
  {Getz}},\ }\href {https://doi.org/10.1038/nature04153} {\bibfield  {journal}
  {\bibinfo  {journal} {Nature}\ }\textbf {\bibinfo {volume} {438}},\ \bibinfo
  {pages} {355} (\bibinfo {year} {2005})}\BibitemShut {NoStop}%
\bibitem [{\citenamefont {Althaus}(2015)}]{althaus_2015}%
  \BibitemOpen
  \bibfield  {author} {\bibinfo {author} {\bibfnamefont {C.~L.}\ \bibnamefont
  {Althaus}},\ }\href {https://doi.org/10.1016/s1473-3099(15)70135-0}
  {\bibfield  {journal} {\bibinfo  {journal} {Lancet Infect. Dis.}\ }\textbf
  {\bibinfo {volume} {15}},\ \bibinfo {pages} {507} (\bibinfo {year}
  {2015})}\BibitemShut {NoStop}%
\bibitem [{\citenamefont {Kucharski}\ and\ \citenamefont
  {Althaus}(2015)}]{kucharski_althaus_2015}%
  \BibitemOpen
  \bibfield  {author} {\bibinfo {author} {\bibfnamefont {A.~J.}\ \bibnamefont
  {Kucharski}}\ and\ \bibinfo {author} {\bibfnamefont {C.~L.}\ \bibnamefont
  {Althaus}},\ }\bibfield  {journal} {\bibinfo  {journal} {Eurosurveillance}\
  }\textbf {\bibinfo {volume} {20}},\ \href
  {https://doi.org/10.2807/1560-7917.es2015.20.25.21167}
  {10.2807/1560-7917.es2015.20.25.21167} (\bibinfo {year} {2015})\BibitemShut
  {NoStop}%
\bibitem [{\citenamefont {Riou}\ and\ \citenamefont
  {Althaus}(2020)}]{riou_althaus_2020}%
  \BibitemOpen
  \bibfield  {author} {\bibinfo {author} {\bibfnamefont {J.}~\bibnamefont
  {Riou}}\ and\ \bibinfo {author} {\bibfnamefont {C.~L.}\ \bibnamefont
  {Althaus}},\ }\href {https://doi.org/10.2807/1560-7917.es.2020.25.4.2000058}
  {\bibfield  {journal} {\bibinfo  {journal} {Eurosurveillance}\ }\textbf
  {\bibinfo {volume} {25}},\ \bibinfo {pages} {2000058} (\bibinfo {year}
  {2020})}\BibitemShut {NoStop}%
\bibitem [{\citenamefont {Althouse}\ \emph {et~al.}(2020)\citenamefont
  {Althouse}, \citenamefont {Wenger}, \citenamefont {Miller}, \citenamefont
  {Scarpino}, \citenamefont {Allard}, \citenamefont {H{\'e}bert-Dufresne},\
  and\ \citenamefont {Hu}}]{althouse2020superspreading}%
  \BibitemOpen
  \bibfield  {author} {\bibinfo {author} {\bibfnamefont {B.~M.}\ \bibnamefont
  {Althouse}}, \bibinfo {author} {\bibfnamefont {E.~A.}\ \bibnamefont
  {Wenger}}, \bibinfo {author} {\bibfnamefont {J.~C.}\ \bibnamefont {Miller}},
  \bibinfo {author} {\bibfnamefont {S.~V.}\ \bibnamefont {Scarpino}}, \bibinfo
  {author} {\bibfnamefont {A.}~\bibnamefont {Allard}}, \bibinfo {author}
  {\bibfnamefont {L.}~\bibnamefont {H{\'e}bert-Dufresne}},\ and\ \bibinfo
  {author} {\bibfnamefont {H.}~\bibnamefont {Hu}},\ }\href
  {https://doi.org/10.1371/journal.pbio.3000897} {\bibfield  {journal}
  {\bibinfo  {journal} {PLoS Bio.}\ }\textbf {\bibinfo {volume} {18}},\
  \bibinfo {pages} {e3000897} (\bibinfo {year} {2020})}\BibitemShut {NoStop}%
\bibitem [{joh(2021)}]{johnshopkins_data}%
  \BibitemOpen
  \href@noop {} {\bibinfo {title} {{COVID-19 Data Repository by the Center for
  Systems Science and Engineering (CSSE) at Johns Hopkins University}}},\
  \bibinfo {howpublished} {\url{https://github.com/CSSEGISandData/COVID-19}}
  (\bibinfo {year} {2021})\BibitemShut {NoStop}%
\bibitem [{\citenamefont {Dong}\ \emph {et~al.}(2020)\citenamefont {Dong},
  \citenamefont {Du},\ and\ \citenamefont {L.}}]{csse_jh_alsocite}%
  \BibitemOpen
  \bibfield  {author} {\bibinfo {author} {\bibfnamefont {E.}~\bibnamefont
  {Dong}}, \bibinfo {author} {\bibfnamefont {H.}~\bibnamefont {Du}},\ and\
  \bibinfo {author} {\bibfnamefont {G.}~\bibnamefont {L.}},\ }\href
  {https://doi.org/10.1016/S1473-3099(20)30120-1} {\bibfield  {journal}
  {\bibinfo  {journal} {Lancet Infect. Dis.}\ }\textbf {\bibinfo {volume}
  {20}},\ \bibinfo {pages} {533} (\bibinfo {year} {2020})}\BibitemShut
  {NoStop}%
\bibitem [{\citenamefont {Allard}\ \emph {et~al.}(2017)\citenamefont {Allard},
  \citenamefont {Althouse}, \citenamefont {Scarpino},\ and\ \citenamefont
  {H{\'e}bert-Dufresne}}]{allard2017asymmetric}%
  \BibitemOpen
  \bibfield  {author} {\bibinfo {author} {\bibfnamefont {A.}~\bibnamefont
  {Allard}}, \bibinfo {author} {\bibfnamefont {B.~M.}\ \bibnamefont
  {Althouse}}, \bibinfo {author} {\bibfnamefont {S.~V.}\ \bibnamefont
  {Scarpino}},\ and\ \bibinfo {author} {\bibfnamefont {L.}~\bibnamefont
  {H{\'e}bert-Dufresne}},\ }\href {https://doi.org/10.1073/pnas.1703073114}
  {\bibfield  {journal} {\bibinfo  {journal} {Proc. Natl. Acad. Sci. U.S.A.}\
  }\textbf {\bibinfo {volume} {114}},\ \bibinfo {pages} {8969} (\bibinfo {year}
  {2017})}\BibitemShut {NoStop}%
\bibitem [{\citenamefont {Vázquez}\ and\ \citenamefont
  {Moreno}(2003)}]{VaquezAlexei2003Rtdo}%
  \BibitemOpen
  \bibfield  {author} {\bibinfo {author} {\bibfnamefont {A.}~\bibnamefont
  {Vázquez}}\ and\ \bibinfo {author} {\bibfnamefont {Y.}~\bibnamefont
  {Moreno}},\ }\href@noop {} {\bibfield  {journal} {\bibinfo  {journal}
  {Physical review. E, Statistical, nonlinear, and soft matter physics}\
  }\textbf {\bibinfo {volume} {67}},\ \bibinfo {pages} {015101} (\bibinfo
  {year} {2003})}\BibitemShut {NoStop}%
\bibitem [{\citenamefont {Hébert-Dufresne}\ \emph {et~al.}(2013)\citenamefont
  {Hébert-Dufresne}, \citenamefont {Allard}, \citenamefont {Young},\ and\
  \citenamefont {Dubé}}]{hebertdufresne2013percolation}%
  \BibitemOpen
  \bibfield  {author} {\bibinfo {author} {\bibfnamefont {L.}~\bibnamefont
  {Hébert-Dufresne}}, \bibinfo {author} {\bibfnamefont {A.}~\bibnamefont
  {Allard}}, \bibinfo {author} {\bibfnamefont {J.-G.}\ \bibnamefont {Young}},\
  and\ \bibinfo {author} {\bibfnamefont {L.~J.}\ \bibnamefont {Dubé}},\
  }\href@noop {} {\bibfield  {journal} {\bibinfo  {journal} {Physical review.
  E, Statistical, nonlinear, and soft matter physics}\ }\textbf {\bibinfo
  {volume} {88}},\ \bibinfo {pages} {062820} (\bibinfo {year}
  {2013})}\BibitemShut {NoStop}%
\bibitem [{\citenamefont {Newman}(2009)}]{NewmanMEJ2009Rgwc}%
  \BibitemOpen
  \bibfield  {author} {\bibinfo {author} {\bibfnamefont {M.~E.~J.}\
  \bibnamefont {Newman}},\ }\href@noop {} {\bibfield  {journal} {\bibinfo
  {journal} {Physical review letters}\ }\textbf {\bibinfo {volume} {103}},\
  \bibinfo {pages} {058701} (\bibinfo {year} {2009})}\BibitemShut {NoStop}%
\bibitem [{\citenamefont {Allard}\ \emph {et~al.}(2012)\citenamefont {Allard},
  \citenamefont {Hébert-Dufresne}, \citenamefont {Noël}, \citenamefont
  {Marceau},\ and\ \citenamefont {Dubé}}]{AllardA2012Bpoa}%
  \BibitemOpen
  \bibfield  {author} {\bibinfo {author} {\bibfnamefont {A.}~\bibnamefont
  {Allard}}, \bibinfo {author} {\bibfnamefont {L.}~\bibnamefont
  {Hébert-Dufresne}}, \bibinfo {author} {\bibfnamefont {P.-A.}\ \bibnamefont
  {Noël}}, \bibinfo {author} {\bibfnamefont {V.}~\bibnamefont {Marceau}},\
  and\ \bibinfo {author} {\bibfnamefont {L.~J.}\ \bibnamefont {Dubé}},\
  }\href@noop {} {\bibfield  {journal} {\bibinfo  {journal} {Journal of
  physics. A, Mathematical and theoretical}\ }\textbf {\bibinfo {volume}
  {45}},\ \bibinfo {pages} {405005} (\bibinfo {year} {2012})}\BibitemShut
  {NoStop}%
\bibitem [{\citenamefont {Karrer}\ and\ \citenamefont
  {Newman}(2010)}]{KarrerBrian2010Rgca}%
  \BibitemOpen
  \bibfield  {author} {\bibinfo {author} {\bibfnamefont {B.}~\bibnamefont
  {Karrer}}\ and\ \bibinfo {author} {\bibfnamefont {M.~E.~J.}\ \bibnamefont
  {Newman}},\ }\href@noop {} {\bibfield  {journal} {\bibinfo  {journal}
  {Physical review. E, Statistical, nonlinear, and soft matter physics}\
  }\textbf {\bibinfo {volume} {82}},\ \bibinfo {pages} {066118} (\bibinfo
  {year} {2010})}\BibitemShut {NoStop}%
\bibitem [{\citenamefont {Allard}\ \emph {et~al.}(2015)\citenamefont {Allard},
  \citenamefont {Hébert-Dufresne}, \citenamefont {Young},\ and\ \citenamefont
  {Dubé}}]{AllardAntoine2015Gaea}%
  \BibitemOpen
  \bibfield  {author} {\bibinfo {author} {\bibfnamefont {A.}~\bibnamefont
  {Allard}}, \bibinfo {author} {\bibfnamefont {L.}~\bibnamefont
  {Hébert-Dufresne}}, \bibinfo {author} {\bibfnamefont {J.-G.}\ \bibnamefont
  {Young}},\ and\ \bibinfo {author} {\bibfnamefont {L.~J.}\ \bibnamefont
  {Dubé}},\ }\href@noop {} {\bibfield  {journal} {\bibinfo  {journal}
  {Physical review. E, Statistical, nonlinear, and soft matter physics}\
  }\textbf {\bibinfo {volume} {92}},\ \bibinfo {pages} {062807} (\bibinfo
  {year} {2015})}\BibitemShut {NoStop}%
\bibitem [{\citenamefont {Melnik}\ \emph {et~al.}(2011)\citenamefont {Melnik},
  \citenamefont {Hackett}, \citenamefont {Porter}, \citenamefont {Mucha},\ and\
  \citenamefont {Gleeson}}]{MelnikSergey2011Tueo}%
  \BibitemOpen
  \bibfield  {author} {\bibinfo {author} {\bibfnamefont {S.}~\bibnamefont
  {Melnik}}, \bibinfo {author} {\bibfnamefont {A.}~\bibnamefont {Hackett}},
  \bibinfo {author} {\bibfnamefont {M.~A.}\ \bibnamefont {Porter}}, \bibinfo
  {author} {\bibfnamefont {P.~J.}\ \bibnamefont {Mucha}},\ and\ \bibinfo
  {author} {\bibfnamefont {J.~P.}\ \bibnamefont {Gleeson}},\ }\href@noop {}
  {\bibfield  {journal} {\bibinfo  {journal} {Physical review. E, Statistical,
  nonlinear, and soft matter physics}\ }\textbf {\bibinfo {volume} {83}},\
  \bibinfo {pages} {036112} (\bibinfo {year} {2011})}\BibitemShut {NoStop}%
\bibitem [{\citenamefont {Rizi}\ \emph {et~al.}()\citenamefont {Rizi},
  \citenamefont {Faqeeh}, \citenamefont {Badie-Modiri},\ and\ \citenamefont
  {Kivelä}}]{RiziAbbasK2021ESaD}%
  \BibitemOpen
  \bibfield  {author} {\bibinfo {author} {\bibfnamefont {A.~K.}\ \bibnamefont
  {Rizi}}, \bibinfo {author} {\bibfnamefont {A.}~\bibnamefont {Faqeeh}},
  \bibinfo {author} {\bibfnamefont {A.}~\bibnamefont {Badie-Modiri}},\ and\
  \bibinfo {author} {\bibfnamefont {M.}~\bibnamefont {Kivelä}},\ }\href@noop
  {} {\bibinfo {title} {Epidemic spreading and digital contact tracing: Effects
  of heterogeneous mixing and quarantine failures}},\ \bibinfo {note}
  {arXiv:2103.12634},\ \Eprint {https://arxiv.org/abs/arXiv:2103.12634}
  {arXiv:arXiv:2103.12634 [physics.soc-ph]} \BibitemShut {NoStop}%
\bibitem [{\citenamefont {Burgio}\ \emph {et~al.}()\citenamefont {Burgio},
  \citenamefont {Steinegger},\ and\ \citenamefont
  {Arenas}}]{BurgioGiulio2021Hits}%
  \BibitemOpen
  \bibfield  {author} {\bibinfo {author} {\bibfnamefont {G.}~\bibnamefont
  {Burgio}}, \bibinfo {author} {\bibfnamefont {B.}~\bibnamefont {Steinegger}},\
  and\ \bibinfo {author} {\bibfnamefont {A.}~\bibnamefont {Arenas}},\
  }\href@noop {} {\bibinfo {title} {Homophily impacts the success of vaccine
  roll-outs}},\ \bibinfo {note} {arXiv:2112.08240},\ \Eprint
  {https://arxiv.org/abs/2112.08240} {arXiv:2112.08240 [physics.soc-ph]}
  \BibitemShut {NoStop}%
\bibitem [{\citenamefont {Hiraoka}\ \emph {et~al.}()\citenamefont {Hiraoka},
  \citenamefont {Rizi}, \citenamefont {Kivelä},\ and\ \citenamefont
  {Saramäki}}]{HiraokaTakayuki2021HIaE}%
  \BibitemOpen
  \bibfield  {author} {\bibinfo {author} {\bibfnamefont {T.}~\bibnamefont
  {Hiraoka}}, \bibinfo {author} {\bibfnamefont {A.~K.}\ \bibnamefont {Rizi}},
  \bibinfo {author} {\bibfnamefont {M.}~\bibnamefont {Kivelä}},\ and\ \bibinfo
  {author} {\bibfnamefont {J.}~\bibnamefont {Saramäki}},\ }\href@noop {}
  {\bibinfo {title} {Herd immunity and epidemic size in networks with
  vaccination homophily}},\ \bibinfo {note} {arXiv:2112.07538},\ \Eprint
  {https://arxiv.org/abs/2112.07538} {arXiv:2112.07538 [physics.soc-ph]}
  \BibitemShut {NoStop}%
\bibitem [{\citenamefont {Cohen}\ \emph {et~al.}(2001)\citenamefont {Cohen},
  \citenamefont {Erez}, \citenamefont {ben Avraham},\ and\ \citenamefont
  {Havlin}}]{PhysRevLett.86.3682}%
  \BibitemOpen
  \bibfield  {author} {\bibinfo {author} {\bibfnamefont {R.}~\bibnamefont
  {Cohen}}, \bibinfo {author} {\bibfnamefont {K.}~\bibnamefont {Erez}},
  \bibinfo {author} {\bibfnamefont {D.}~\bibnamefont {ben Avraham}},\ and\
  \bibinfo {author} {\bibfnamefont {S.}~\bibnamefont {Havlin}},\ }\href
  {https://doi.org/10.1103/PhysRevLett.86.3682} {\bibfield  {journal} {\bibinfo
   {journal} {Phys. Rev. Lett.}\ }\textbf {\bibinfo {volume} {86}},\ \bibinfo
  {pages} {3682} (\bibinfo {year} {2001})}\BibitemShut {NoStop}%
\bibitem [{\citenamefont {H\'ebert-Dufresne}\ and\ \citenamefont
  {Allard}(2019)}]{PhysRevResearch.1.013009}%
  \BibitemOpen
  \bibfield  {author} {\bibinfo {author} {\bibfnamefont {L.}~\bibnamefont
  {H\'ebert-Dufresne}}\ and\ \bibinfo {author} {\bibfnamefont {A.}~\bibnamefont
  {Allard}},\ }\href {https://doi.org/10.1103/PhysRevResearch.1.013009}
  {\bibfield  {journal} {\bibinfo  {journal} {Phys. Rev. Research}\ }\textbf
  {\bibinfo {volume} {1}},\ \bibinfo {pages} {013009} (\bibinfo {year}
  {2019})}\BibitemShut {NoStop}%
\end{thebibliography}
\end{document}